\documentclass[10pt,conference,screen,review]{IEEEtran}
\IEEEoverridecommandlockouts

\usepackage{cite}
\usepackage{xspace}
\usepackage{colortbl}
\usepackage{xcolor}
\usepackage{multirow}
\usepackage{pifont}
\usepackage{makecell}
\usepackage{hyperref}
\usepackage[T1]{fontenc}
\usepackage[scaled=0.85]{beramono}
\usepackage{listings}
\usepackage{xargs} 
\usepackage{array}
\usepackage{makecell}
\usepackage{amsmath}
\usepackage{xspace}
\usepackage{amsthm}
\usepackage[most]{tcolorbox}
\usepackage{amsthm}
\usepackage{enumitem}
\usepackage{soul}
\usepackage[switch]{lineno}
\usepackage[ruled,linesnumbered]{algorithm2e}
\usepackage[normalem]{ulem}
\usepackage{tabularx}
\usepackage{flushend}
\usepackage{rotating}
\usepackage{booktabs}
\usepackage{subcaption}
\usepackage{adjustbox}  % For scaling tables

\def\BibTeX{{\rm B\kern-.05em{\sc i\kern-.025em b}\kern-.08em
    T\kern-.1667em\lower.7ex\hbox{E}\kern-.125emX}}

\definecolor{LightGray}{rgb}{0.97,0.97,0.97}
\definecolor{DarkGray}{rgb}{0.25,0.25,0.25}

\usepackage[colorinlistoftodos,prependcaption,textsize=small]{todonotes}

\usepackage{newfloat}
\DeclareFloatingEnvironment[
  fileext=lol,
  name=Listing
]{listing}

\definecolor{pblue}{rgb}{0.13,0.13,1}
\definecolor{pgreen}{rgb}{0,0.5,0}
\definecolor{pred}{rgb}{0.9,0,0}
\definecolor{pgrey}{rgb}{0.46,0.45,0.48}
\definecolor{backcolor}{rgb}{0.95,0.95,0.92}
\definecolor{darkpurple}{rgb}{0.39, 0.10, 0.40}
\definecolor{darkyellow}{rgb}{0.7, 0.6, 0.1}

\newcommand{\RQ}[1]{RQ\textsubscript{#1}}

\newcommand\digitstyle{\color{orange}}
\makeatletter
\newcommand{\ProcessDigit}[1]
{%
  \ifnum\lst@mode=\lst@Pmode\relax%
   {\digitstyle #1}%
  \else
    #1%
  \fi
}
\makeatother
\lstdefinelanguage{YAML}{
  literate=
    {0}{{{\ProcessDigit{0}}}}1
    {1}{{{\ProcessDigit{1}}}}1
    {2}{{{\ProcessDigit{2}}}}1
    {3}{{{\ProcessDigit{3}}}}1
    {4}{{{\ProcessDigit{4}}}}1
    {5}{{{\ProcessDigit{5}}}}1
    {6}{{{\ProcessDigit{6}}}}1
    {7}{{{\ProcessDigit{7}}}}1
    {8}{{{\ProcessDigit{8}}}}1
    {9}{{{\ProcessDigit{9}}}}1
    {<=}{{\(\leq\)}}1,
  tabsize = 2,
  showstringspaces=false,
  morecomment=[l][\color{gray}]{\#},      
  morecomment=[l][\color{pgreen}]{\'},
  morecomment=[l][\color{blue}]{...},
  morestring=[b][\color{OliveGreen}]{\"},  
  alsoletter=!?-,
  keywords=[1]{
    name, description, type, enum, required, features, parameters, default, configuration, ordering_parameters, ordering_parameters, conjunctive_filters, disjoint_filters, complete_filters, all_value, parameter, values, produces, operationId, get, paths, responses, \$ref, schema, in, x-example, maximum
  },
  keywordstyle=[1]\color{red},
  keywords=[2]{false, true},
  keywordstyle=[2]\color{orange},
  moredelim=**[is][\color{red}]{@}{@},
  moredelim=**[is][\color{black}]{?}{?},
  moredelim=**[is][\color{pgreen}]{^}{^}
}

\lstdefinelanguage{Decls}{
alsoletter=!?-:,
keywords=[1]{
    ppt, :::ENTER, ppt-type, :::EXIT, :::EXIT1, :::EXIT2, :::EXIT3, :::EXIT4, :::EXIT16, variable, decl-version
  },
  keywordstyle=[1]\color{pblue},
keywords=[2]{
    var-kind, dec-type, rep-type, enclosing-var, array
  },
  keywordstyle=[2]\color{pred},
  moredelim=[is][\color{black}]{?}{?},
  morestring=[b][\color{pgreen}]{\"},
  keywords=[3]{false, true},
  keywordstyle=[3]\color{darkyellow},
  moredelim=[is][\color{orange}]{@}{@}
}

\lstdefinestyle{listingtop}{
  float=tp,
  floatplacement=tbp,
  belowcaptionskip=-0.5cm,
}

\newcommand{\approach}{SATORI\xspace}
\newcommand{\baseline}{AGORA+\textsubscript{U}\xspace}
\newcommand{\postmanTool}{PostmanAssertify\xspace}
\newcommand{\dataset}{OKAMI\xspace}
\newcommand{\datasetMeaning}{Oracle Knowledge of API Methods for Innovation\xspace}

\definecolor{lightred}{RGB}{255, 204, 204}
\definecolor{lightgreen}{RGB}{204, 255, 204}
\definecolor{slightlyDarkBlue}{RGB}{135, 206, 235}
\definecolor{lightblue}{RGB}{224, 235, 255}
\definecolor{lightyellow}{RGB}{255, 255, 204}
\definecolor{lightpurple}{RGB}{210, 190, 255}

\makeatletter
\newcommand{\linebreakand}{%
  \end{@IEEEauthorhalign}
  \hfill\mbox{}\par
  \mbox{}\hfill\begin{@IEEEauthorhalign}
}
\makeatother

\begin{document}

% \title{\fontsize{23pt}{28pt}\selectfont \approach: Static Test Oracle Generation for REST APIs}
% \title{\approach: Static Test Oracle Generation for REST APIs\vspace{-1cm}}
\title{\approach: Static Test Oracle Generation\\for REST APIs}


% \author{Anonymous authors}

\author{\IEEEauthorblockN{Juan C. Alonso}
\IEEEauthorblockA{\textit{SCORE Lab, I3US Institute} \\
\textit{Universidad de Sevilla}\\
Seville, Spain \\
javalenzuela@us.es}
\and
\IEEEauthorblockN{Alberto Martin-Lopez}
\IEEEauthorblockA{\textit{SEART @ Software Institute} \\
\textit{Universit\`a della Svizzera italiana}\\
Lugano, Switzerland \\
alberto.martin@usi.ch}
\and
\IEEEauthorblockN{Sergio Segura}
\IEEEauthorblockA{\textit{SCORE Lab, I3US Institute} \\
\textit{Universidad de Sevilla}\\
Seville, Spain \\
sergiosegura@us.es}
\linebreakand
\IEEEauthorblockN{Gabriele Bavota}
\IEEEauthorblockA{\textit{SEART @ Software Institute} \\
\textit{Universit\`a della Svizzera italiana}\\
Lugano, Switzerland\\
gabriele.bavota@usi.ch}
\and
\IEEEauthorblockN{Antonio Ruiz-Cort\'es}
\IEEEauthorblockA{\textit{SCORE Lab, I3US Institute} \\
\textit{Universidad de Sevilla}\\
Seville, Spain \\
aruiz@us.es}
}

\maketitle


\lstset{
  language=YAML,
  basicstyle=\fontsize{6.3pt}{5.8pt}\ttfamily,    
  backgroundcolor=\color{LightGray},
  columns=fullflexible,
  breaklines=false,
  sensitive=true,
  % --------------------------
  frame=bt,
  % --------------------------
  numbers=left,
  numberstyle=\tiny\color{pgrey},
  numbersep=9pt,
  tabsize=2,
  showtabs=false,
  xleftmargin=15pt,
  xrightmargin=5pt,
  frame=single,
  framesep=3pt,
  literate={á}{{\'a}}1
  {ó}{{\'o}}1
}

\definecolor{pblue}{rgb}{0.13,0.13,1}
\definecolor{pgreen}{rgb}{0,0.5,0}
\definecolor{pred}{rgb}{0.9,0,0}
\definecolor{darkyellow}{rgb}{0.7, 0.6, 0.1}
\lstdefinelanguage{Json}{
moredelim=**[is][\color{pred}]{@}{@},
moredelim=**[is][\color{pblue}]{*}{*},
moredelim=**[is][\color{pgreen}]{\^}{\^},
moredelim=**[is][\color{darkyellow}]{ñ}{ñ}
}

\begin{abstract}
%REST API test case generation tools have increasingly become more sophisticated, handling complex inputs with increasing automation. However, despite their strengths in input generation, these tools are constrained by the types of test oracles they support, often limited to crashes, regressions, and non-compliance with API specifications or design standards.
%This article introduces \approach (Static API Test ORacle Inference), a black-box approach for generating test oracles for REST APIs by analyzing their OpenAPI Specification (the industry standard). \approach uses large language models to infer the expected behavior of an API by analyzing the properties of the response fields of its operations, such as their name and descriptions. We extended the \postmanTool tool to automatically convert the test oracles reported by \approach into executable assertions. \approach, despite being a static approach, outperformed the state-of-the-art dynamic approach (which requires executing the API to generate test oracles), with an F1-Score of 74.3\% vs. 69.3\% on 17 operations from 12 industrial APIs. Also, our extensive evaluation shows that static and dynamic oracle inference are complementary. \approach uncovered 18 bugs in popular APIs---Amadeus Hotel, Deutschebahn, FDIC, GitLab, Marvel, OMDb and Vimeo---leading to bug fixes and documentation updates.

REST API test case generation tools are evolving rapidly, with growing capabilities for the automated generation of complex tests. However, despite their strengths in test data generation, these tools are constrained by the types of test oracles they support, often limited to crashes, regressions, and non-compliance with API specifications or design standards.
This paper introduces \approach (Static API Test ORacle Inference), a black-box approach for generating test oracles for REST APIs by analyzing their OpenAPI Specification. \approach uses large language models to infer the expected behavior of an API by analyzing the properties of the response fields of its operations, such as their name and descriptions. To foster its adoption, we extended the \postmanTool tool to automatically convert the test oracles reported by \approach into executable assertions. Evaluation results on 17 operations from 12 industrial APIs show that \approach can automatically generate up to hundreds of valid test oracles per operation. \approach achieved an F1-score of 74.3\%, outperforming the state-of-the-art dynamic approach AGORA+ (69.3\%)---which requires executing the API---when generating comparable oracle types. Moreover, our findings show that static and dynamic oracle inference methods are complementary: together, \approach and AGORA+ found 90\% of the oracles in our annotated ground-truth dataset. Notably, \approach uncovered 18 bugs in popular APIs (Amadeus Hotel, Deutschebahn, FDIC, GitLab, Marvel, OMDb and Vimeo) leading to documentation updates by the API maintainers.
\end{abstract}

\begin{IEEEkeywords}
REST APIs, test oracle, LLM, automated testing
\end{IEEEkeywords}

\section{Introduction}
\label{sec:introduction}

Web Application Programming Interfaces (APIs) allow heterogeneous software systems to communicate over the network~\cite{jacobson11-book,richardson13-book}. Among these, REST APIs---those adhering to the REpresentational State Transfer (REST) architectural style~\cite{fielding00-phd}---have become the predominant standard. REST APIs organize their functionality around distinct resources (e.g., a video in the Vimeo API~\cite{vimeo-api}) that clients access and manipulate through HTTP interactions. REST APIs underpin the business models of major companies such as Google, Microsoft, and Uber~\cite{jacobson11-book}. 
The Postman 2024 State of the API Report~\cite{postman2024StateAPIReport} shows that APIs are crucial business assets, with 62\% of developers working on revenue-generating APIs.

The importance of REST APIs has led to the development of numerous techniques and tools for automated test case generation for these systems~\cite{golmohammadi2022Survey,Myeongsoo2022ISSTA}. Most techniques follow a black-box approach, deriving test cases automatically from the OpenAPI Specification (OAS)~\cite{oai} of the API under test. These test cases are generated by assigning values to the input parameters and validating the returned responses using various test oracles~\cite{barr2015Oracle}, which serve as mechanisms for determining whether the output of a program is correct for a given input. Despite their promising results in generating valid API requests, these tools are all limited by the types of failures they can detect, primarily crashes (5XX HTTP status code responses)~\cite{restler2019,arcuri2019restful,alberto20icsoc,Karlsson2020,Viglianisi2020,hatfield2021deriving,wu2022Combinatorial}, disconformities with the API specification (e.g., an undocumented output JSON property)~\cite{arcuri2019restful,Viglianisi2020,Karlsson2020,alberto20icsoc,hatfield2021deriving}, regressions~\cite{godefroid2020Issta}, and violations of API best practices (e.g., ensuring that repeated calls to idempotent operations return identical responses)~\cite{Atlidakis2020}. For example, 
% Listing~\ref{lst:sampleResponse} shows a response for the ``getBusinesses'' operation of the Yelp API which conforms to the API specification (Listing~\ref{lst:oas}) and would be considered correct by existing tools. 
given the API specification of the ``getBusinesses'' operation of the Yelp API shown in Listing~\ref{lst:oas}, Listing~\ref{lst:sampleResponse} shows an API response which conforms to such specification and would be considered correct by existing tools.
However, this response may still contain errors that would go undetected by test case generators, such as incorrect field length (e.g., \texttt{country} should have 2 characters), format (e.g., \texttt{image\_url} should be a valid URL), or violations of numerical constraints (e.g., \texttt{latitude} should range from -90 to 90), among others. Recent surveys~\cite{golmohammadi2022Survey} and tool comparisons~\cite{martinFSE22Online,Myeongsoo2022ISSTA} highlight test oracle generation as a key challenge in automated test case creation for REST APIs. This is the problem that motivates our work.

To the best of our knowledge, the only existing approach in the literature that addresses the automated generation of test oracles for REST APIs is AGORA+~\cite{Alonso2023AGORA,Alonso2025AGORATosem}, which generates test oracles through the detection of likely invariants (i.e., properties of the output that should always hold). Invariants are detected by analyzing the API specification and a set of API requests with their corresponding responses. Although effective, the main limitation of AGORA+ is its reliance on a sufficiently diverse test suite that thoroughly exercises the API functionality to report accurate invariants. If the test suite lacks diversity or contains faulty responses, the reported invariants may be wrong or incomplete.

This paper presents \approach (Static API Test ORacle Inference), a black-box static approach for automatically generating test oracles for REST APIs by analyzing their OAS document, without requiring prior API execution. \approach leverages large language models (LLMs) to infer test oracles from the unstructured components of the OAS document, such as response field names and descriptions, making it compatible with existing API testing tools that support OAS. Currently, \approach supports a catalog of 17 types of test oracles, which can be easily extended. To foster its adoption, we extended the \postmanTool tool~\cite{Alonso2025AGORATosem} to transform the test oracles reported by \approach into executable JavaScript assertions, written using the Chai library~\cite{chaiAssertionLibrary}, that are compatible with Postman~\cite{postmanAPIPlatform}, a widely used API platform in industry with over 40 million users.

The results of an evaluation conducted on 17 operations from 12 industrial APIs show the capabilities of \approach to automatically generate up to hundreds of valid test oracles per API operation, achieving an F1-Score of 74.3\%, better than AGORA+ (69.3\%) in generating the same types of oracles supported by both approaches. Moreover, \approach identified 18 real bugs across 7 widely used industrial APIs (vs. 13 bugs in 7 APIs by AGORA+), which would have passed unnoticed by existing test case generators. Our findings led to documentation updates in the API of Vimeo. Since it does not require prior API execution, \approach offers a more cost-effective solution than AGORA+. Our thorough evaluation also shows that each approach excels in identifying distinct oracle types, making them complementary: the combination of \approach and AGORA+ found 90\% of the test oracles of an annotated ground-truth dataset.

This paper
% first outlines the background and related work on testing REST APIs and test oracle generation techniques (Section~\ref{sec:background}). Then, it presents
makes the following research and engineering contributions:

% \vspace{-0.8mm}
\begin{itemize}
\item \approach, a black-box static approach for automatically generating test oracles for REST APIs through specification analysis.
% \approach is open-source and publicly available on GitHub~\cite{satoriRepository}.

% \item A carefully crafted prompt which can be reused with any arbitrary LLM to generate test oracles for REST APIs.

\item \dataset, a dataset containing the annotated ground truth of all the test oracles of the API operations used in our evaluation (over 10.5k test oracles from more than 1.8k response fields), enabling benchmarking and future comparisons. OKAMI is publicly available on Hugging Face~\cite{okamiDataset}.

\item An extension of \postmanTool~\cite{Alonso2025AGORATosem} that transforms the test oracles generated by \approach into executable JavaScript assertions compatible with the widely used Postman API platform~\cite{postmanAPIPlatform}.
% This tool is open-source and publicly available on GitHub~\cite{postmanAssertifyStatic}.% \aruiz{I think that the current version of the paper does not devote the space expected for a principal contribution.}

\item An assessment of 21 LLMs as the backbone of \approach, compared in terms of size, coding and reasoning capabilities, and cost.

\item An empirical evaluation of \approach and AGORA+ in terms of precision, recall, F1-Score and failure detection across 17 operations from 12 industrial APIs, including the discovery of 22 real-world bugs.

% \item A publicly available replication package including the source code, and the data used in our work, as well as a pre-configured virtual machine to ease reproducibility and replicability~\cite{}.
\end{itemize}
% \vspace{-0.8mm}

% Section~\ref{sec:threats} discusses threats to validity, and Section~\ref{sec:conclusions} concludes the paper.

All our code and data are publicly available~\cite{satoriRepository}.

\section{Background and Related Work}
\label{sec:background}

\subsection{Automated Testing of REST APIs}

Web APIs commonly adhere to the REpresentational State Transfer (REST)~\cite{fielding00-phd} architectural style, being known as REST APIs~\cite{richardson13-book}. REST APIs typically comprise multiple RESTful web services, each implementing CRUD (create, read, update, delete) operations on a resource (e.g., in the Vimeo API~\cite{vimeo-api}, a resource is a video). These operations are usually invoked by sending HTTP requests (generally \texttt{GET}, \texttt{POST}, \texttt{PUT} and \texttt{DELETE}) to a Uniform Resource Identifier (URI) representing a resource or a collection of resources.

\begin{listing}
\begin{lstlisting}[language=YAML,caption=OAS excerpt of the Yelp API., label=lst:oas, captionpos=b, frame=tlrb,belowskip=-4.1mm]
paths:
  '/businesses/search':
    get:
      operationId: getBusinesses
      parameters:
       - name: term
          description: 'Search term, e.g. food or restaurants.'
          in: query
          schema:
            type: string
       - name: location
          description: 'Geographic area for business search.'
          in: query
          schema:
            type: string
      responses:
        '200':
          description: 'Returns all businesses'
          @content@:
            @application/json@:
              schema:
                type: object
                @properties@:
                  @total@:
                    type: integer
                    description: 'Total number of businesses found.'
                  @businesses@:
                    type: array
                    @items@:
                      type: object
                      @properties@:
                        @id@:
                          type: string
                        name:
                          type: string
                        @image_url@:
                          type: string
                        @rating@:
                          type: number
                          description: 'Business rating (ranges from 1 ... 5).'
                        @coordinates@:
                          type: object
                          @properties@:
                            @latitude@:
                              type: number
                            @longitude@:
                              type: number
                        @price@:
                          type: string
                          description: 'Price level. Value is
                          ^one of $, $$, $$$ and $$$$.^'
                          @example@: '$$'
                        @location@:
                          type: object
                          @properties@:
                            @city@:
                              type: string
                            @country@:
                              type: string
                              description: 'ISO 3166-1 alpha-2 country code.'
\end{lstlisting}
% \vspace{-0.5cm}
\end{listing}

\begin{listing}%[htpb]
\begin{lstlisting}[language=Json, caption={Yelp API response in JSON format.}, label={lst:sampleResponse}, captionpos=b, belowskip=-2.0mm]
{
    *"total"*: ^1^,
    *"businesses"*: [
        {
            *"id"*: @"7dzGDH1BtzEjhZh1FeeaqA"@,
            *"name"*: @"Caipirinha Corner"@,
            *"image_url"*: @"https://s3-media1.fl.yelpcdn.com/bphoto/zrG.jpg"@,
            *"rating"*: ^4.0^,
            *"coordinates"*: {
                *"latitude"*: ^37.3968404980258^,
                *"longitude"*: ^-5.97877264022827^
            },
            *"price"*: @"$"@,
            *"location"*: {
                *"city"*: @"Seville"@,
                *"country"*: @"ES"@
            }
        }
    ]
}
\end{lstlisting}
\vspace{-0.5cm}
\end{listing}

REST APIs are commonly described using the OpenAPI Specification (OAS)~\cite{oai} format, arguably the industry standard. An OAS document outlines the API operations, detailing their input parameters and responses. For instance, Listing~\ref{lst:oas} shows an excerpt from the OAS of the ``getBusinesses'' operation of the Yelp API~\cite{yelp-api}. The specification defines the HTTP method and URI required to call the operation (lines 1–3), operation ID (line 4), input parameters (lines 5–15), and response format (lines 16–60). Listing~\ref{lst:sampleResponse} shows an API response aligning with this specification.

Automated testing of REST APIs often employs a black-box approach~\cite{MartinLopez2021Restest,Viglianisi2020,restler2019,hatfield2021deriving,alonso22TSE,Karlsson2020,segura2018metamorphic,Atlidakis2020,Godefroid2020Intelligent,godefroid2020Issta,wu2022Combinatorial,liu2022Morest,NLPtoREST,EDEFuzz,kimASE2023,kim2024leveraging,yandrapallyIcse23,ase24Deeprest,icst24Kat}, where, based on an OAS, these methods generate pseudo-random test cases (sequences of HTTP requests) and test oracles (assertions on the responses). Techniques vary in how they generate API calls (i.e., test inputs), leveraging methods like property-based testing~\cite{Karlsson2020,hatfield2021deriving,segura2018metamorphic,segura2022Automated}, model-based testing~\cite{MartinLopez2021Restest,liu2022Morest}, and constraint-based testing~\cite{alberto20icsoc,tsc-martinlopez20,icst24Kat}. Some approaches target individual API operations and create single API requests, while others design sequences of API calls for stateful testing~\cite{restler2019,Viglianisi2020,hatfield2021deriving,kimASE2023,ase24Deeprest,icst24Kat}. White-box approaches, which require access to the API source code, are less common, and most existing techniques use search algorithms to maximize failure detection and code coverage~\cite{stallenberg2022Improving,arcuri2019restful}.
Recent approaches for API testing leverage LLMs~\cite{kim2024leveraging,icst24Kat,kim2024AutoRestTest,kim2025LlamaRestTest} and reinforcement learning~\cite{kimASE2023,ase24Deeprest} to extract realistic input values and dependencies between parameters and operations, but none of them tackle the oracle problem.

Generated test oracles for failure detection primarily target API crashes (e.g., 5XX status codes) and API specification violations~\cite{alberto20icsoc,Viglianisi2020,hatfield2021deriving,arcuri2019restful,Karlsson2020}, with some also addressing regressions~\cite{godefroid2020Issta} and design practices~\cite{Atlidakis2020}. However, these approaches are limited in identifying issues beyond syntax, overlooking domain-specific assertions like those in Listing~\ref{lst:sampleResponse}. For example, they miss validations such as ensuring that the \texttt{country} response field value has two characters, \texttt{latitude} and \texttt{longitude} fall within specific ranges, or \texttt{price} adheres to allowed values (``\$'', ``\$\$'', ``\$\$\$'', ``\$\$\$\$'').

Some approaches infer input or output constraints in REST APIs through static analysis~\cite{grent2021automatically,huang2024generating}. These constraints can be considered as test oracles in the form of pre- and post-conditions. However, these approaches require the source code of the system, which may not always be available (as in the case of the industrial APIs tested in our work) and thus cannot operate in black-box mode. More importantly, they derive constraints based on the \emph{implemented} behavior, which may be faulty, thus limiting the usefulness of the inferred constraints.

\sloppy{To the best of our knowledge, the only approach for inferring domain-specific oracles for REST APIs is AGORA+~\cite{Alonso2023AGORA,Alonso2025AGORATosem}, which uses invariant detection to generate test oracles. Invariants are output properties that should always hold (e.g., \texttt{LENGTH(return.location.country)==2}), and they are detected by analyzing patterns in previous API executions (i.e., request/response pairs). The effectiveness of AGORA+ depends on having a sufficiently diverse test suite: if the suite lacks variety or includes faulty responses, the detected invariants may be incomplete or invalid. However, many of these oracles can be inferred directly from the response field information in the OAS (e.g., response field names and descriptions, as shown in Listing~\ref{lst:oas}), without prior API execution. This is the goal of \approach.}

\subsection{Test Oracle Generation}
Automated techniques for generating test oracles can be categorized by their inputs and the domains they target. Oracles can be derived from source code~\cite{mastropaolo2021Studying,yu2022Automated}, formal specifications~\cite{gay2014Improving}, semi-structured documentation~\cite{zhai2020Translating}, previous program executions~\cite{molina2022icse,Chen2021Boosting,kapugama2022Human,ibrahimzada2022Perfect,douIcse23}, or combinations of these inputs. Application contexts include databases~\cite{douIcse23}, Java programs~\cite{molina2022icse}, cyber-physical systems~\cite{ayerdi2021Generating}, and machine learning programs~\cite{Houssem2020TestingML}, among others.

Other related techniques include metamorphic testing, regression testing and invariant detection. Metamorphic testing~\cite{segura2018metamorphic,segura2016Survey,segura2022Automated,EDEFuzz} uses manually identified relationships between inputs and outputs across multiple executions of the system under test. Regression testing~\cite{fraser2013Whole} compares the observed behavior to previous software versions to verify that changes do not disrupt existing features. Invariant detection identifies properties expected to consistently hold in program outputs, which can serve as test oracles to verify the correctness of outputs. They can be detected either statically, by analyzing code (without executing it)~\cite{cousot92Static}, or dynamically, by examining program behavior across executions~\cite{ernst07Daikon,molina2022icse,Alonso2023AGORA,Alonso2025AGORATosem}.

In recent years, LLMs have been applied across various stages of the software testing lifecycle~\cite{tse24SurveyLLMsInTesting}, including unit test case generation~\cite{schafer24TseEmpirical,Yuan24fseEvaluating,Alshahwan24FseAutomated,Bhatia24llm4code}, test input generation~\cite{xia24IcseFuzz4All,sunase23SMT}, debugging~\cite{feng24IcsePrompting,li24AseNuances}, and program repair~\cite{gao24AseWhatMakes}. 
LLM-based techniques proposed for tackling the oracle problem~\cite{hossain2024togllcorrectstrongtest,yu2022Automated,fse23NeuralBased,fse24Empirical,fse24CanLLMs,ast22Generating,molinelli2025tratto} focus on specific programming languages and operate at the method level, leveraging information such as variable names and dataflow analysis to infer test oracles, thus making them unsuitable for the domain of REST APIs.
To the best of our knowledge, \approach is the first approach to leverage LLMs for addressing the oracle problem specifically in black-box testing of REST APIs.

\section{\approach}
\label{sec:approach}

\begin{figure}[h]
  \centering
  \includegraphics[width=1.0\linewidth]{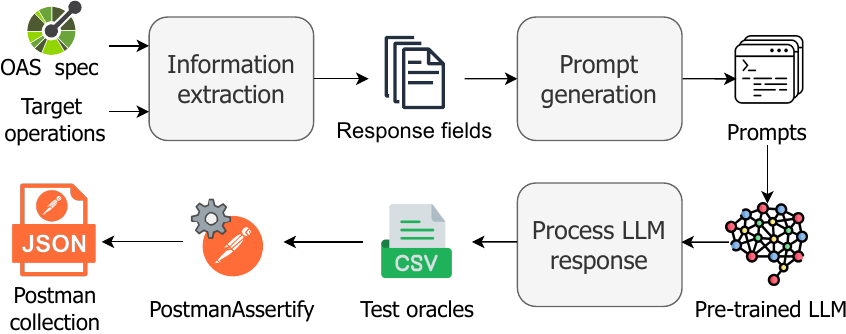}
  \vspace{-5mm}
  \caption{Workflow of \approach.}
  \label{fig:approach}
\end{figure}

Figure~\ref{fig:approach} outlines \approach, our approach for automatically generating test oracles for REST APIs through specification analysis. Starting from an OAS, \approach extracts the schema of each response field of all the target operations and generates prompts for a (configurable) LLM. The outputs of the LLM (i.e., test oracles) are then processed into a machine-readable format and, following an optional human verification step, are provided to an extended version of \postmanTool~\cite{satoriRepository} to produce a Postman collection with executable test assertions.
% \aruiz{Creo que habría que valorar si denominar este workflow SATORI(<nombre de LLM>) e incluso delinear sus límites de modo que se queden fuera sólo las entradas y las salidas. con el layout actual sólo es pintar una caja. De otro modo, si se quiere ilustrar los experimentos realizados, si sería necesario indicar que hay N LLMs y que se han probado N OAS.  Por otra parte, quizás el LLM habría que meterlo dentro de una caja azul que entiendo representa a una actividad e ilustrar la respuesta del LLM que posteriormente es procesada.}
% \aruiz{the figure does not reflect that the model is chosen by the user}

We now describe the complete \approach workflow in detail, as well as the supported set of test oracles.

\subsection{Automated Response Field Prompt Generation}
This subsection explains how \approach generates prompts to infer test oracles for each response field of an API operation.

\subsubsection{Information Extraction}

First, we extract contextual information to generate input prompts. Specifically, we gather details for each response field---such as name, description, and examples---along with global context like the API name and operation ID, providing the LLM with richer context to infer accurate test oracles.

\subsubsection{Prompt Generation}

Using information from the previous step, \approach generates one prompt per response field. This prompt has been designed following well-established prompting techniques and patterns~\cite{whitePlop23PromptPattern,schulhoff2025promptreportsystematicsurvey}.
% The left-hand side of the table displays the prompt template, while the right-hand side provides an example for the \texttt{price} response field from Listing~\ref{lst:oas}. 
We use a \emph{System prompt} to set the overall behavior of the model and ensure consistent outputs. Then, the main prompt begins by establishing context for the task, followed by the necessary information, and then a detailed task description. Consequently, each prompt is structured into three sections: \emph{Context prompt}, \emph{Properties prompt}, and \emph{Oracles prompt}. Our supplemental material~\cite{satoriRepository} contains all the prompts generated by \approach for our evaluation. In what follows, we provide an example of each prompt section, highlighting in boldface the dynamic parts of the prompt.

\paragraph{System Prompt} We use a role-playing approach, instructing the model to act as an expert software engineer.
% The system prompt is:

\begin{tcolorbox}[colback=LightGray, colframe=black, sharp corners=southwest, boxrule=0.8pt, left=2mm, right=0mm, top=0mm, bottom=0mm, enhanced, borderline west={4pt}{0pt}{black}]
\small
\emph{\noindent
You are a highly skilled software engineer with extensive experience in designing and testing REST APIs. Answer to your questions simply by generating a JSON object, without providing any additional information or explanation.
}
\end{tcolorbox}

\paragraph{Context Prompt} This provides the model with essential context, including the name of the API, the operation under test, and the name and type of the target response field.
% Its value is:

\begin{tcolorbox}[colback=LightGray, colframe=black, sharp corners=southwest, boxrule=0.8pt, left=2mm, right=0mm, top=0mm, bottom=0mm, enhanced, borderline west={4pt}{0pt}{black}]
\small
\emph{\noindent
I am going to give you a response field of the \textbf{getBusinesses} operation of the \textbf{Yelp} API.
The name of this response field is ``\textbf{price}'' and it is of type \textbf{string}.
}
\end{tcolorbox}

\paragraph{Properties Prompt} This includes all additional properties of the response field available in the API specification, such as descriptions or examples, which the LLM can analyze to identify potential test oracles.
% For instance, its value for the \texttt{price} response field of Listing~\ref{lst:oas} is:

\begin{tcolorbox}[colback=LightGray, colframe=black, sharp corners=southwest, boxrule=0.8pt, left=2mm, right=0mm, top=0mm, bottom=0mm, enhanced, borderline west={4pt}{0pt}{black}]
\small
\emph{\noindent
This response field has the following properties:\\
\textbf{``name'': ``price''\\
``type'': ``string''\\
``description'': ``Price level. Value is one of \$, \$\$, \$\$\$, \$\$\$\$.''\\
``example'': ``\$\$''}
}
\end{tcolorbox}

\paragraph{Oracles Prompt} 
This final section guides the model to generate test oracles for the response field and is structured in three parts. First, the \emph{Task introduction prompt}
% (value: \emph{``Given this information, I want you to answer the following questions about some
% properties of this response field:''})
outlines the task. Next, several \emph{Single oracle prompts} are presented as questions, guiding the model to infer specific test oracles based on the datatype of the response field (Section~\ref{sec:targetOracles}). Each single oracle prompt consists of a question and specifies the expected response as a JSON property name and datatype, or a default JSON property value if no oracle is identified.
% For instance, for the \texttt{string\_specific\_values} test oracle, \approach generates the following single oracle prompt: \emph{``3 - Should this response field have a set of specific values? JSON property:
% 'string\_specific\_values', of type array of string, if there are no specific values, the
% array is empty''}
Finally, the \emph{Response format prompt} instructs the model to return the test oracles in a structured JSON.

\begin{tcolorbox}[colback=LightGray, colframe=black, sharp corners=southwest, boxrule=0.8pt, left=2mm, right=0mm, top=0mm, bottom=0mm, enhanced, borderline west={4pt}{0pt}{black}, breakable]
\small
\textbf{\colorbox{black}{\textcolor{white}{Task introduction prompt}}}\vspace{1mm}

\emph{\noindent
Given this information, I want you to answer the following questions about some properties of this response field:\\
}
- - - - - - - - - - - - - - - - - - - - - - - - - - - - - - - - - - - - - - - - \\
\textbf{\colorbox{black}{\textcolor{white}{Single oracle prompts (one example)}}}\vspace{1mm}

\emph{\noindent
\textbf{3} - \textbf{Should this response field have a set of specific values?} JSON property:\\
``\textbf{string\_specific\_values}'', of type \textbf{array of string}, \textbf{if there are no specific values, the array is empty}\\
}
- - - - - - - - - - - - - - - - - - - - - - - - - - - - - - - - - - - - - - - - \\
\textbf{\colorbox{black}{\textcolor{white}{Response format prompt}}}\vspace{1mm}

\emph{\noindent
I want the response to be a single JSON object with~the properties indicated in each question (\textbf{string\_is\_url, string\_is\_numeric, string\_specific\_values, string\_is\_email, string\_is\_date, string\_fixed\_length, string\_is\_time}). I don't want any kind of additional natural language explanation, only the JSON~object.
}
\end{tcolorbox}

% \aruiz{yo indicaría el tamaño del prompt, creo que deja más completa la sección y deja claro si está por debajo de los 8K que creo es el tamaño de algunos de los modelos más pequeños. tb estaría bien conocer el número de tokens en el que cada LLM fragmenta el prompt.}

\subsection{LLM Response Processing}
\approach processes the responses of the LLM to ensure syntactic correctness, handling issues like transforming responses into valid JSON, standardizing formats, merging multiple JSONs, and removing spurious text.
Listing~\ref{lst:satoriResponse} shows the response returned by \approach for the \texttt{price} response field.

\begin{listing}[h]
\vspace{-2mm}
\centering
\noindent\begin{minipage}{\linewidth}
\begin{lstlisting}[language=Json, caption={Example of test oracles generated by SATORI.}, label={lst:satoriResponse}, captionpos=b, belowskip=-2.0mm]
{
   *"string_is_url"*: ñfalseñ,
   *"string_is_numeric"*: ñfalseñ,
   *"string_specific_values"*: [ @"$"@, @"$$"@, @"$$$"@, @"$$$$"@ ],
   *"string_is_email"*: ñfalseñ,
   *"string_is_date"*: ñfalseñ,
   *"string_fixed_length"*: ^null^,
   *"string_is_time"*: ñfalseñ
}
\end{lstlisting}
\end{minipage}
% \vspace{-0.5cm}
\end{listing}

% Processed responses are then consolidated into a single CSV file, where each line represents a response field and its corresponding test oracles.
% This CSV file, along with the OAS specification, serves as input to an enhanced version of \postmanTool~\cite{postmanAssertifyStatic}. \postmanTool converts the test oracles into executable JavaScript assertions compatible with Postman~\cite{postmanAPIPlatform}, a popular platform for API development and testing. Postman enables users to send HTTP requests to REST APIs and analyze responses within an intuitive graphical interface. It organizes API requests into collections, facilitating the execution of entire test suites at once. \postmanTool produces a Postman collection that includes sample API requests for each one of the tested operations, attaching the test oracles to each request as executable assertions. For instance, \postmanTool would generate the assertion: \texttt{pm.expect(["\$", "\$\$", "\$\$\$", "\$\$\$\$"].includes(price)).to.be.true} for the response field in Table~\ref{tab:mappingTable}.

These test oracles, together with the OAS document, are provided as input to our \postmanTool extension, which transforms them into executable JavaScript assertions compatible with Postman~\cite{postmanAPIPlatform}. \postmanTool produces a Postman collection of API requests for each tested operation, embedding the test oracles in each request and thus making the approach readily applicable in practice. For instance, the generated test cases could be executed programmatically, via the Postman GUI, or integrated into CI/CD pipelines. An example of a generated assertion is \texttt{pm.expect(["\$", "\$\$", "\$\$\$", "\$\$\$\$"].includes(price)).to.be.true}, corresponding to the \texttt{price} response field (line 4 of Listing~\ref{lst:satoriResponse}).

\begin{table}[b]
\vspace{-2mm}
\centering
\caption{Test oracles supported by \approach.}
\vspace{-2mm}
\resizebox{\columnwidth}{!}{
  \begin{tabular}{l|p{0.75\columnwidth}}
    \toprule
    \textbf{Datatype} & \textbf{Test oracles} \\
    \hline
    String & \texttt{is\_url}, \texttt{is\_numeric}, \texttt{specific\_values}, \texttt{is\_email}, \\
    & \texttt{is\_date}, \texttt{fixed\_length}, \texttt{is\_time} \\
    \hline
    Boolean & \texttt{always\_true}, \texttt{always\_false} \\
    \hline
    Number & \texttt{min\_value}, \texttt{max\_value}, \texttt{specific\_values} \\
    \hline
    Array & \{String,Boolean,Number\}-oracles, \texttt{min\_size}, \texttt{max\_size}, \\
    & \texttt{specific\_sizes} \\
    \hline
    Array{[}number{]} & \{Array\}-oracles, \texttt{asc\_order}, \texttt{desc\_order} \\
    \bottomrule
    \end{tabular}
}
\label{tab:satoriOracles}
% \vspace{-5mm}
\end{table}

\subsection{Target Oracles}
\label{sec:targetOracles}
\approach supports a set of 17 types of test oracles, shown in Table~\ref{tab:satoriOracles}. Note that string, boolean and number oracles can be applied to elements of arrays (fourth row of Table~\ref{tab:satoriOracles}). These oracles support all 49 unary invariants (i.e., test oracles evaluating a single variable) supported by the dynamic approach AGORA+~\cite{Alonso2023AGORA,Alonso2025AGORATosem}, which were derived from a systematic study of the oracles found in 40 real-world APIs.
We focus on unary oracles to make our evaluation affordable, since deriving a ground-truth dataset of $n$-ary oracles would lead to a combinatorial explosion, requiring the manual annotation of a dataset of up to tens of thousands of instances per API operation (see Section~\ref{sec:exp1-setup}).
Our proposed oracles are simpler than AGORA+'s while supporting the same use cases. For instance, float and integer invariants of AGORA+ (e.g., \texttt{OneOfFloat} and \texttt{OneOfScalar}) are combined into a single \approach test oracle (e.g., \texttt{number\_specific\_values}). The resulting oracles assess properties such as string formats (e.g., \texttt{string\_is\_url}), numerical boundaries (e.g., \texttt{number\_max\_value}), and ordering of arrays (e.g., \texttt{array\_number\_asc\_order}). We refer the reader to the \approach documentation~\cite{satoriRepository} for a complete list of the supported test oracles. These can be extended to support specific requirements.

\section{Evaluation}
\label{sec:evaluation}

We aim to answer the following research questions:

% \textbf{\RQ1:} \emph{What is the effectiveness of \approach in generating test oracles?} We aim to measure the performance of \approach in generating test oracles that model the expected API behavior.

% \textbf{\RQ2:} \emph{What are the failure detection capabilities of the test oracles generated by \approach?}
% The final goal is generating test oracles that can be used to identify erroneous responses. Thus, we aim to investigate the effectiveness of the reported oracles for detecting failures in API responses.

\textbf{\RQ1:} \emph{How do different LLMs perform in generating test oracles with \approach?} We analyze the performance and cost of different LLMs as the backbone of \approach, considering model size, code specialization and reasoning capabilities.

\textbf{\RQ2:} \emph{What is the effectiveness of \approach in generating test oracles and how does it compare against dynamic oracle generation approaches?} We compare the oracle generation capabilities of \approach equipped with the LLM chosen in the previous RQ with respect to AGORA+ as a representative dynamic approach.

\textbf{\RQ3:} \emph{How effective is \approach in detecting artificially seeded faults and how does it compare against dynamic approaches?}
We evaluate the effectiveness of the oracles generated by \approach in detecting faults (mutations) in API responses, comparing it against AGORA+.

\textbf{\RQ4:} \emph{How effective is \approach in detecting real faults?} We evaluate the ability of \approach to detect real faults in API responses, especially those not identified by AGORA+.

\textbf{\RQ5:} \emph{How much does it cost to find a bug with \approach? Can this cost be saved?} As GPT-4o is the most effective model with \approach, we compute the cost per bug found (in dollars) and investigate whether free open-source models can find the same bugs.

\subsection{Experiment 1: Test Oracle Generation}
\label{sec:experiment1}

With this experiment, we aim to answer \RQ1 and \RQ2 by measuring the performance achieved by \approach with different LLMs, and comparing it against dynamic oracle generation approaches.

\subsubsection{Experimental Setup}
\label{sec:exp1-setup}
Next we describe the dataset used as a benchmark, the LLMs and baselines experimented with, and the metrics considered for evaluation.

\paragraph{Dataset}
As a contribution of this work, we present the \dataset (\datasetMeaning) dataset~\cite{okamiDataset}, a reliable benchmark for evaluating test oracle generation techniques for REST APIs. \dataset is composed of 17 operations from 12 industrial APIs, which were used for the evaluation of the oracle generation approach AGORA+~\cite{Alonso2023AGORA,datasetAGORA,Alonso2025AGORATosem} and in previous papers~\cite{alonso22TSE,martinFSE22Online,restler2019}.
When necessary, we updated the OAS documents of these APIs according to the latest version of the web docs.  We manually created the ground truth of all the oracles supported by \approach for all the response fields of these API operations (e.g., labeling \texttt{href} response fields as URLs).
Due to the extremely costly effort of manually annotating thousands of response fields, we randomly sampled API operations from the AGORA+ benchmark until having annotated, at least, 10k oracles. This resulted in a dataset of 17 API operations, 1,816 response fields and 10,645 test oracles.
% The labeling process was carried out by the first author. To minimize human bias or errors, the second author analyzed a subset of 50 randomly sampled response fields, accounting for \textcolor{red}{XXX} test oracles. No conflicts were found in the analysis of such subset by both annotators.
To avoid human bias or errors during the labeling process, we carefully analyzed the API specification (OAS) for labeling each response field and consulted the API providers in case of doubts or discrepancies.
\dataset is publicly available on Hugging Face~\cite{okamiDataset} and as part of our supplemental material~\cite{satoriRepository} to serve as a benchmark for future studies.

\paragraph{LLMs and Baselines}
To answer \RQ1, we selected a set of 21 LLMs according to different criteria, namely: (i)~model size (from 1B parameters to hundreds of billions of closed-source models), (ii)~code specialization (explicitly trained on code or not), and (iii)~reasoning capabilities (explicitly trained to reason about their answers or not). We selected models from six different vendors, i.e., Google, Meta, Microsoft, Alibaba, DeepSeek and OpenAI. We aim to analyze the impact of the aforementioned criteria on the performance of \approach, and to select the best model for the subsequent experiments.

Regarding configuration parameters, we use the default settings for all models and a temperature of 0 (greedy decoding), thus making their outputs mostly deterministic.

To answer \RQ2, we compare the performance of \approach against AGORA+, a dynamic approach that requires prior API execution to infer test oracles. In particular, we consider two versions of AGORA+: unary and binary. The unary version, denoted as \baseline, reports the same test oracles as \approach (i.e., those involving a single variable), while the binary version, denoted as AGORA+, generates also test oracles involving two variables (e.g., \texttt{input.limit >= size(return.items[])}).
% The \dataset dataset contains only unary oracles, as creating the ground truth for all the possible binary oracles would lead to a combinatorial explosion, requiring the manual annotation of a dataset of tens of thousands of instances per API operation. For this reason, for \baseline we report precision, recall and F1-Score, overall and per oracle type, while for AGORA+ we report only the overall precision.

The authors of AGORA+ explain the need for a sufficiently diverse set of API requests and responses to detect invariants effectively, with 50 being enough. For a fair evaluation, we used the same sets of 10k requests used in the AGORA+ paper~\cite{Alonso2025AGORATosem}. Since the performance of AGORA+ depends on these (randomly generated) sets of request-response pairs, the authors selected 10 subsets of 50 pairs each from among these 10k instances, and computed averages. We used the same 10 sets in this work.

% All these details seem too specific for readers to understand, and they don't belong in this section, therefore I would omit them:
% Optional properties within an OAS response field, such as \texttt{enum} or \texttt{regex}, can help indicate the presence of potential test oracles. We followed a conservative approach and removed these properties when providing the response fields as inputs to \approach. There were 174 response fields (9.6\%) containing these machine-readable properties in the whole dataset, with 150 of them belonging to two APIs (GitHub and YouTube).
% In some OAS specifications (GitHub, Spotify and YouTube), a response schema is reused across multiple operations, and some of its response fields are never present in real API responses of certain operations (for example, the GitHub operations used in our evaluation never return any of the response fields of the template repository), and hence AGORA+ is unable to report oracles for them.
% For a fair comparison, we omitted these absent response fields from GitHub (311), Spotify (16) and YouTube (45). In YouTube, some of the response fields are private data that is only shown to the video owner (such as video processing details) that would require uploading videos for AGORA+ to generate test oracles. We omitted these response fields for a fair comparison.

\captionsetup[subfigure]{font=footnotesize}
\begin{figure*}[t]
    \centering
    \begin{subfigure}[b]{0.306\textwidth}
        \includegraphics[width=\textwidth]{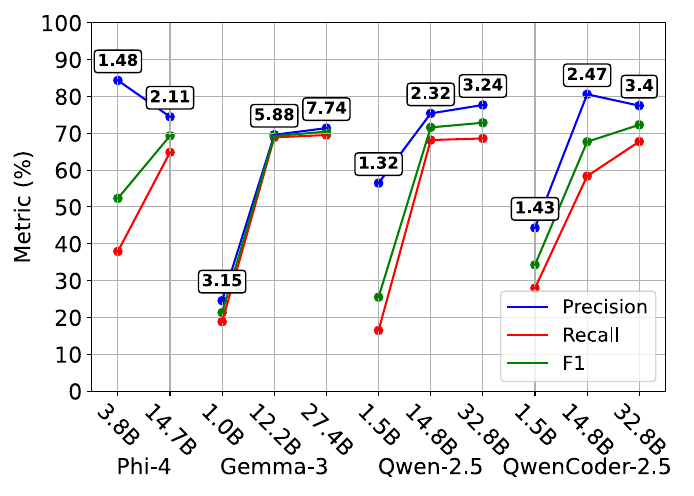}
        \caption{Size.}
        \label{fig:rq1a}
    \end{subfigure}
    \hfill
    \begin{subfigure}[b]{0.217\textwidth}
        \includegraphics[width=\textwidth]{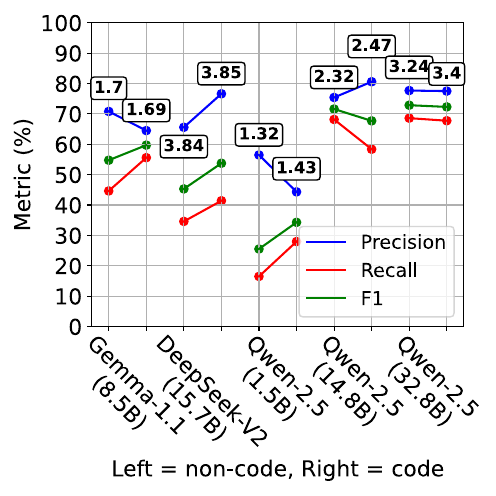}
        \caption{Code specialization.}
        \label{fig:rq1b}
    \end{subfigure}
    \hfill
    \begin{subfigure}[b]{0.195\textwidth}
        \includegraphics[width=\textwidth]{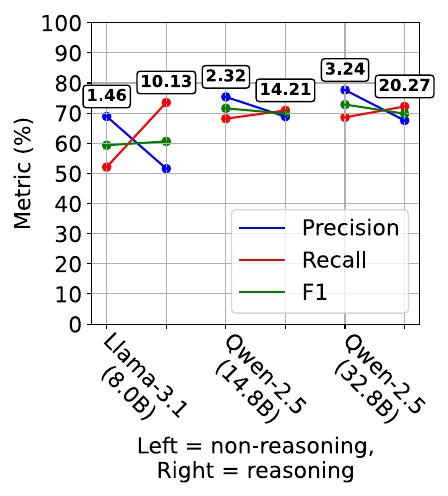}
        \caption{Reasoning capabilities.}
        \label{fig:rq1c}
    \end{subfigure}
    \hfill
    \begin{subfigure}[b]{0.262\textwidth}
        \includegraphics[width=\textwidth]{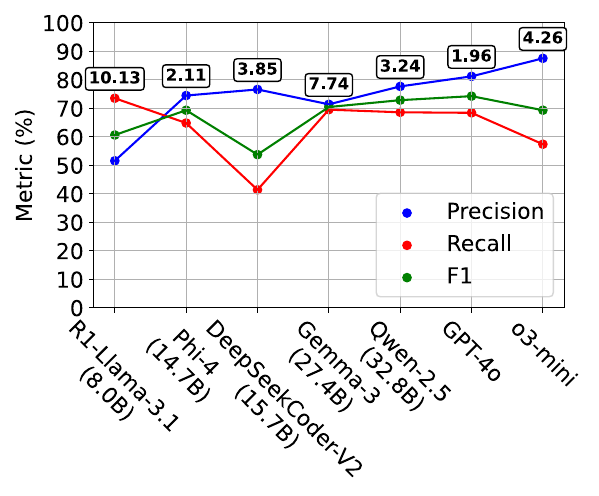}
        \caption{Best model of each vendor.}
        \label{fig:rq1d}
    \end{subfigure}
    \caption{\RQ1: Precision, recall and F1-Score of each model evaluated according to several criteria. Numbers above markers denote average time (in seconds) to generate oracles for a single response field.}
    \label{fig:rq1}
\vspace{-4mm}
\end{figure*}

\paragraph{Metrics}
For both \RQ1 and \RQ2, we report the overall precision, recall and F1-Score. Here precision refers to the percentage of correct oracles generated by a technique $T_i$ out of the total number of oracles generated by $T_i$. Recall, instead, captures the percentage of oracles in our ground truth dataset that has been generated by $T_i$.  For \RQ2, we report also the same metrics per oracle type (see 17 types of oracles in Table \ref{tab:satoriOracles}) for further analyses. For AGORA+ (binary), we report only the overall precision, since the \dataset dataset contains only unary oracles.
% This is because creating the ground truth for all the possible binary oracles would lead to a combinatorial explosion, requiring the manual annotation of a dataset of up to tens of thousands of instances per API operation.
We also report the average time (in seconds) required to generate the test oracles for each response field of the API operations. The open-source LLMs were executed on a single NVIDIA A100 GPU with 80GB of VRAM, while OpenAI models were invoked via their web API~\cite{openai-api}. AGORA+ does not require GPU resources, therefore it was executed on a desktop computer equipped with an Intel i9-12900K @3.20GHz, 64GB RAM, and 2TB SSD running Windows 11.

% The metrics reported refer to a single execution for LLMs and to the average of 10 executions for AGORA+. For open source LLMs we used greedy decoding, while for OpenAI models we set the temperature to 0, thus turning the randomness in the models' outputs negligible. On the other hand, the effectiveness 

% \begin{figure*}[t]
%   \centering
%   \includegraphics[width=1.0\textwidth]{Figures/plots_rq1.pdf}
%   \caption{RQ1.}
%   \label{fig:rq1}
% \end{figure*}

\subsubsection{\RQ1: Experimental Results}
Figure~\ref{fig:rq1} shows the precision, recall and F1-Score achieved by each model.
% It also shows the average time (in seconds) for each model to generate all possible test oracles for a single response field in the OAS.
The figure is split in four subfigures according to the criteria previously mentioned, i.e., size (\ref{fig:rq1a}), code specialization (\ref{fig:rq1b}), reasoning capabilities (\ref{fig:rq1c}) and best model of each vendor considered (\ref{fig:rq1d}).
The numbers shown on top of markers denote the average time (in seconds) to generate all possible oracles of a single response field in the OAS.

As expected, model size plays a role, as confirmed in Figure~\ref{fig:rq1a}. Here we evaluated four families featuring the same model in different sizes, namely, Phi-4 (3.8B, 14.7B), Gemma 3 (1B, 12.2B, 27.4B), Qwen2.5 and Qwen2.5-Coder (1.5B, 14.8B, 32.8B). As observed, models under 4B parameters exhibit significantly lower performance compared to the rest. However, models between 12-15B parameters achieve performance (69.3--71.6\%) mostly on par with that achieved by models with $\sim$30B parameters (70.4--72.2\%). This may be relevant if cutting costs is desirable (e.g., cheaper GPUs and faster inference times).

To evaluate the impact of code specialization on the task of oracle generation for REST APIs, we evaluated five models which offer a base version and a \emph{code-specialized} version (i.e., the same model further trained on code), namely, Gemma 1.1 (8.5B), DeepSeek-V2-Lite (15.7B) and Qwen2.5 (1.5B, 14.8B, 32.8B). As shown in Figure~\ref{fig:rq1b}, results are mixed. While code specialization seems to help for Gemma, DeepSeek and Qwen 1.5B models (7.6\% higher F1 on average), it has a slightly negative impact on Qwen 14.8B (-3.9\% F1) and no significant impact on Qwen 32.8B.
% This suggests that code specialization may not be critical for this task, especially for medium-size general-purpose models which already achieve good results.
Execution times are not significantly affected by code specialization.

Regarding reasoning capabilities, we evaluated three models which offer versions distilled from (i.e., fine-tuned with the answers generated by) the reasoning model DeepSeek R1~\cite{guo2025deepseekR1}, namely, Llama 3.1 (8B) and Qwen2.5 (14.8B, 32.8B). Figure~\ref{fig:rq1c} highlights two interesting aspects. First, reasoning distillation does not significantly affect the overall F1-Score of models, although it worsens precision and improves recall, meaning that distilled models tend to generate more oracles, resulting in more false positives (wrong oracles) but also less false negatives (less correct oracles missed). On the other hand, reasoning models are significantly slower than their non-distilled counterparts, taking 6-7$\times$ longer to generate oracles.

Figure~\ref{fig:rq1d} shows the comparison between the best model of each vendor and the closed source models GPT-4o and o3-mini. The average F1-Score ranges from 53.7\% (DeepSeekCoder-V2-Lite) up to 74.3\% (GPT-4o). The low performance of DeepSeekCoder-V2-Lite for this task may be attributed to the fact that is a \emph{Mixture of Experts} (MOE) model originally designed with over 200B parameters, thus its lite version may not be able to unleash the full potential of the MOE architecture.

\begin{tcolorbox}[colback=LightGray, title=Answer to \RQ1: Comparison of LLMs, colframe=black, left=1mm, right=1mm, top=1mm, bottom=1mm]
% The best LLM for SATORI is GPT-4o, with an average F1-Score of \textcolor{red}{74.3\%} and the fastest execution time (1.96s per response field). Other cost-effective alternatives are Phi-4 (14.7B parameters, 69.3\% F1-Score, 2.11s) and Qwen2.5 (32.8B parameters, 72.8\% F1-Score, 3.24s).
Models under 4B parameters exhibit significantly lower performance ($<$55\% F1-Score) compared to their larger counterparts ($\sim$70\% F1-Score). Code specialization helps in some cases, while reasoning distillation does not affect the F1-Score, but increases execution times. Overall, the best model for SATORI is GPT-4o (74.3\% F1-Score, 1.96s execution time).
\end{tcolorbox}

\begin{figure*}[t]
  \centering
  \begin{subfigure}[l]{0.5\textwidth}
    \includegraphics[width=\textwidth]{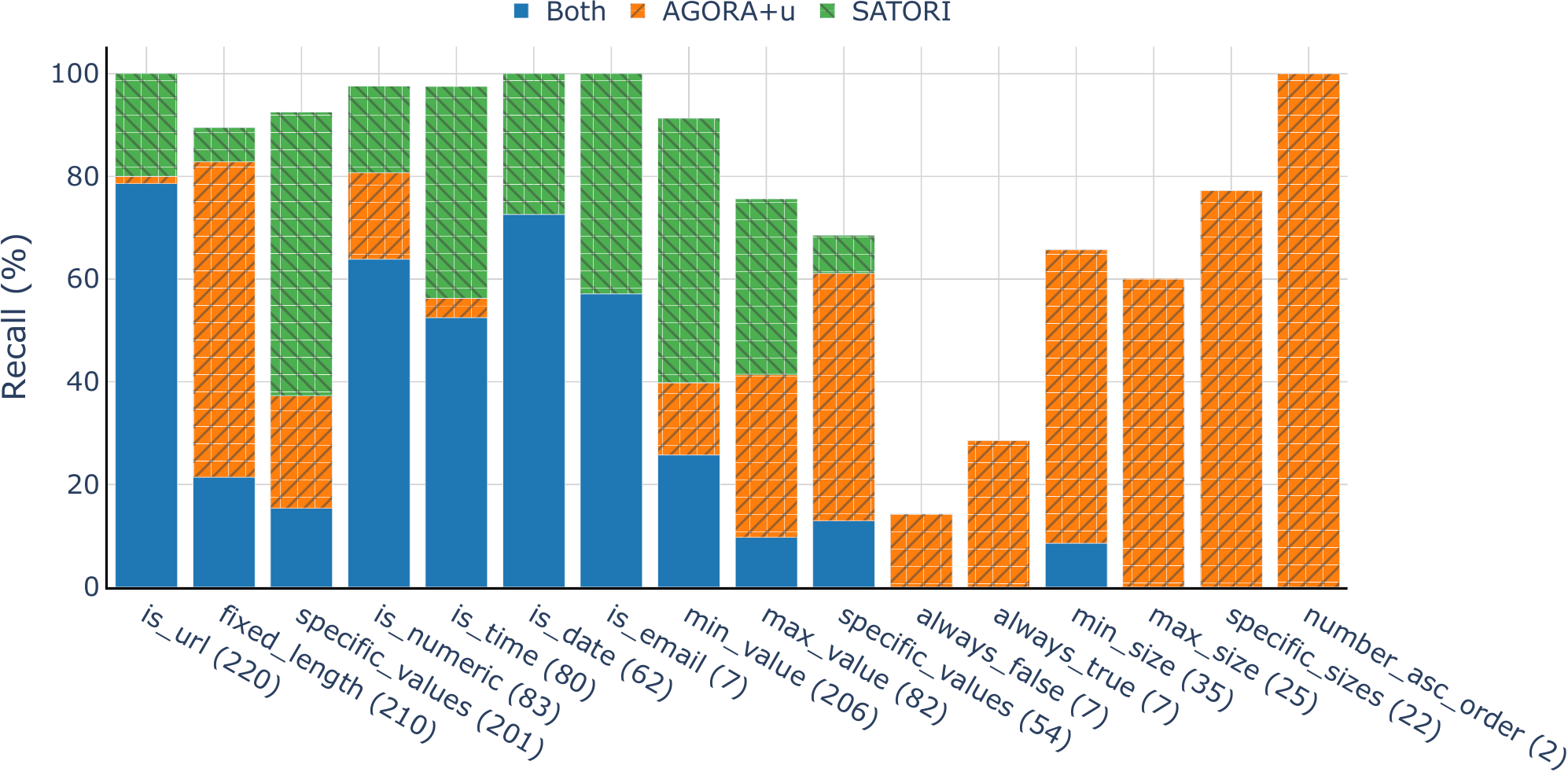}
    % \vspace{-8mm}
    \caption{Per oracle type.}
    \label{fig:oracleOverlapping}
  \end{subfigure}
  \hspace{-5mm}
  \begin{subfigure}[r]{0.5\textwidth}
    \vspace{0.5mm}
    \includegraphics[width=\textwidth]{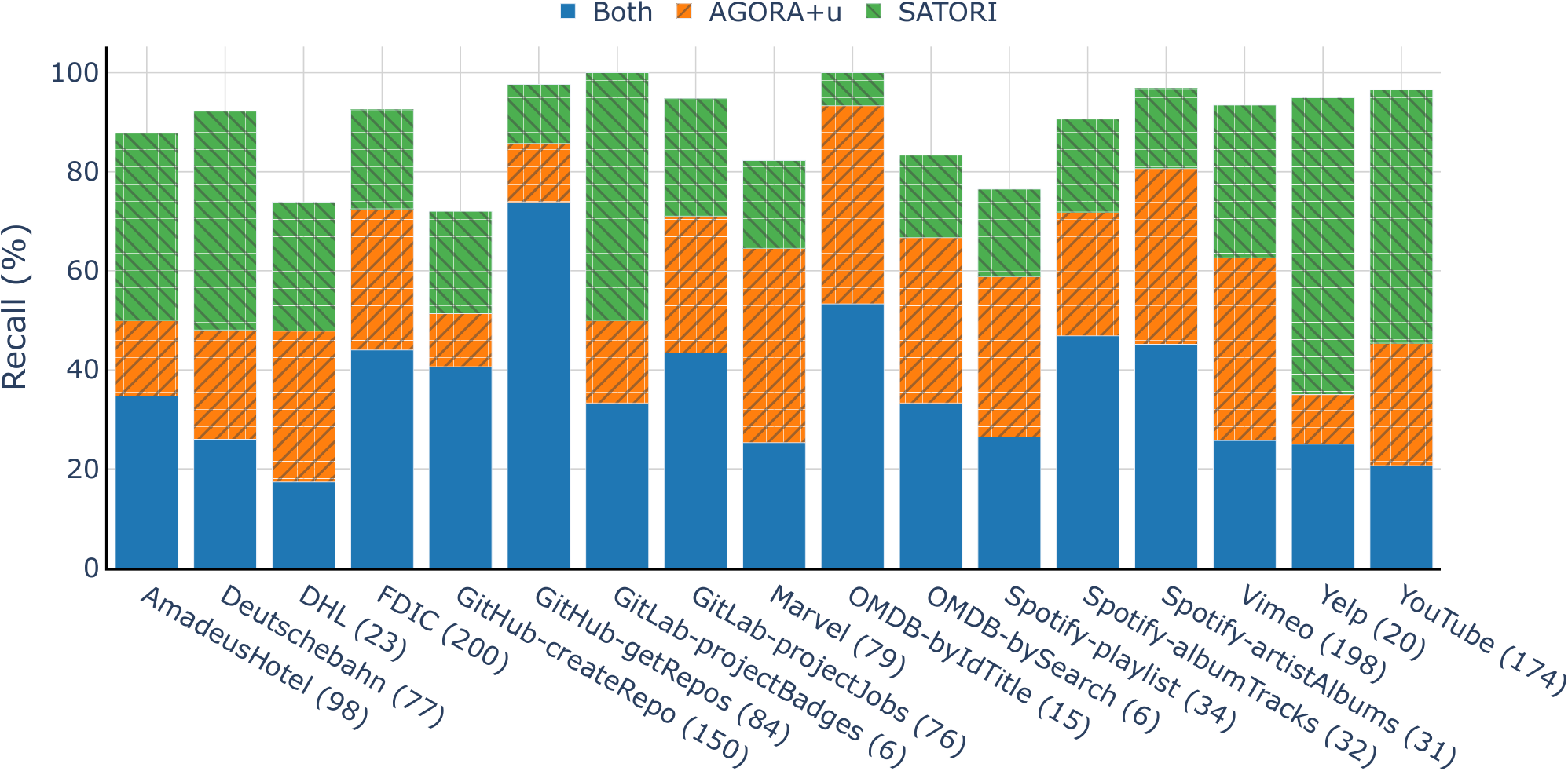}
    \vspace{-4.5mm}
    \caption{Per API operation.}
    \label{fig:operationOracleOverlapping}
  \end{subfigure}
  \caption{\RQ2: Overlapping of recall (percentage of oracles detected) between \approach and \baseline.}
  \label{fig:combinedOverlapping}
\vspace{-3mm}
\end{figure*}

\subsubsection{\RQ2: Experimental Results}

% Table~\ref{tab:experimentalResultsRQ2} shows the performance, in terms of precision (P), recall (R) and F1-Score (F1), for each type of test oracle achieved by \approach (equipped with GPT4o) and by \baseline.
% For each test oracle, we report the count of response fields of its datatype that are positive instances (the oracle is present) and negative instances (the oracle is absent).
% Table~\ref{tab:experimentalResultsRQ2} combines primitive and array test oracles. For instance, \texttt{string\_is\_url} and \texttt{array\_string\_is\_url} are reported as \texttt{string\_is\_url}. Table~\ref{tab:agoraBinaryResults} presents the precision achieved by AGORA+ on each operation, with an overall precision of 56\%, ranging from 29.5\% to 100\%.

Table~\ref{tab:experimentalResultsRQ2} shows the performance, in terms of precision (P), recall (R) and F1-Score (F1), as well as true and false positives and negatives (TP, TN, FP, FN) for each type of test oracle and overall achieved by \approach (equipped with GPT-4o) and \baseline. The table combines oracles related to primitive and array datatypes (e.g., \texttt{string\_is\_url} and \texttt{array\_string\_is\_url} are both considered \texttt{string\_is\_url}). The table does not show oracles not found in the ground-truth dataset and for which no approach generated false positives (i.e., \texttt{array\_number\_desc\_order}).

The results show that \baseline achieved higher F1-Score than \approach for 9 out of 16 types of oracles, while \approach is better for the remaining 7. Even so, the overall F1-Score of \approach (74.3\%) remains higher than that of \baseline (69.3\%). For the binary version of AGORA+, the precision of AGORA+ is 68.8\%, significantly lower than the precision of both \approach (81.2\%) and \baseline (76.8\%), meaning that AGORA+ tends to generate significantly more false positives.

\begin{table}[t]
\vspace{1.5mm}
\centering
\caption{\RQ2: Test oracle generation by \approach and \baseline, per oracle type and overall.}
\resizebox{1.0\linewidth}{!}{
{
\large
\setlength{\tabcolsep}{3pt}
\begin{tabular}{c|l|rrrrrrr|rrrrrrr}
\toprule
\multicolumn{2}{c}{} & \multicolumn{7}{c}{\textbf{\baseline}} & \multicolumn{7}{c}{\textbf{\approach (GPT-4o)}} \\
\cmidrule(l){3-9} \cmidrule(l){10-16}
\multicolumn{2}{c}{\multirow{-2}{*}{\textbf{Type \& Oracle}}} & \multicolumn{1}{c}{\textbf{P}} & \multicolumn{1}{c}{\textbf{R}}  & \multicolumn{1}{c}{\textbf{F1}} & \multicolumn{1}{c}{\textbf{TP}} & \multicolumn{1}{c}{\textbf{TN}} & \multicolumn{1}{c}{\textbf{FP}} & \multicolumn{1}{c}{\textbf{FN}} & \multicolumn{1}{c}{\textbf{P}} & \multicolumn{1}{c}{\textbf{R}}  & \multicolumn{1}{c}{\textbf{F1}} & \multicolumn{1}{c}{\textbf{TP}} & \multicolumn{1}{c}{\textbf{TN}} & \multicolumn{1}{c}{\textbf{FP}} & \multicolumn{1}{c}{\textbf{FN}} \\
\midrule
\multirow{7}{*}{\rotatebox{90}{String}}
& is\_url & 99.9 & 80 & 88.9 & 176.1 & 803.8 & 0.2 & 43.9 & 94.3 & 98.6 & \textbf{\underline{96.4}} & 217 & 791 & 13 & 3 \\
& fixed\_length & 81.9 & 83.6 & \textbf{\underline{82.7}} & 173.9 & 777.4 & 38.6 & 34.1 & 59.6 & 28.6 & 38.7 & 59 & 778 & 40 & 147 \\
& specific\_values & 47.6 & 42.5 & 44.9 & 74.5 & 766.6 & 82 & 100.9 & 65.1 & 85 & \textbf{\underline{73.8}} & 142 & 781 & 76 & 25 \\
& is\_numeric & 99.1 & 77.8 & \textbf{\underline{87.2}} & 64.6 & 940.4 & 0.6 & 18.4 & 84.8 & 80.7 & 82.7 & 67 & 929 & 12 & 16 \\
& is\_time & 100 & 55 & 70.9 & 44 & 944 & 0 & 36 & 97.4 & 93.8 & \textbf{\underline{95.5}} & 75 & 942 & 2 & 5 \\
& is\_date & 100 & 72.9 & 84.3 & 45.2 & 962 & 0 & 16.8 & 95.4 & 100 & \textbf{\underline{97.6}} & 62 & 959 & 3 & 0 \\
& is\_email & 100 & 42.9 & 59 & 3 & 1017 & 0 & 4 & 100 & 100 & \textbf{\underline{100}} & 7 & 1017 & 0 & 0 \\
\midrule
\multirow{3}{*}{\rotatebox{90}{Number}}
& min\_value & 81.4 & 43.6 & 56.7 & 82.5 & 23.7 & 18.9 & 106.9 & 82 & 91.9 & \textbf{\underline{86.6}} & 159 & 24 & 35 & 14 \\
& max\_value & 66.2 & 44.2 & 52.9 & 34 & 137.5 & 17.5 & 43 & 90 & 44.4 & \textbf{\underline{59.5}} & 36 & 147 & 4 & 45 \\
& specific\_values & 65.6 & 66.3 & \textbf{\underline{65.9}} & 33.7 & 163.4 & 17.8 & 17.1 & 78.6 & 20.4 & 32.4 & 11 & 175 & 3 & 43 \\
\midrule
\multirow{2}{*}{\rotatebox{90}{Bool}}
& always\_false & 10.5 & 14.3 & \textbf{\underline{12}} & 1 & 117.3 & 8.7 & 6 & - & 0 & - & 0 & 126 & 0 & 7 \\
& always\_true & 35.2 & 27.1 & \textbf{\underline{30.3}} & 1.9 & 122.3 & 3.7 & 5.1 & - & 0 & - & 0 & 126 & 0 & 7 \\
\midrule
\multirow{4}{*}{\rotatebox{90}{Array}}
& min\_size & 58.4 & 63.5 & \textbf{\underline{60.8}} & 22.3 & 64.9 & 16 & 12.8 & 100 & 8.6 & 15.8 & 3 & 81 & 0 & 32 \\
& max\_size & 46.3 & 59.2 & \textbf{\underline{51.9}} & 14.2 & 75.4 & 16.6 & 9.8 & 0 & 0 & - & 0 & 91 & 3 & 22 \\
& specific\_sizes & 47.8 & 77.1 & \textbf{\underline{59}} & 16.2 & 77.2 & 17.8 & 4.8 & 0 & 0 & - & 0 & 94 & 3 & 19 \\
& number\_asc\_order & 100 & 100 & \textbf{\underline{100}} & 2 & 1 & 0 & 0 & - & 0 & - & 0 & 1 & 0 & 2 \\
\midrule
\multicolumn{1}{l}{} & \multicolumn{1}{l}{\textbf{TOTAL}} & \multicolumn{1}{c}{\textbf{76.8}} & \multicolumn{1}{c}{\textbf{63.2}} & \multicolumn{1}{c}{\textbf{69.3}} & \multicolumn{1}{c}{\textbf{789.1}} & \multicolumn{1}{c}{\textbf{6993.9}} & \multicolumn{1}{c}{\textbf{238.4}} & \multicolumn{1}{c|}{\textbf{459.6}} & \multicolumn{1}{c}{\textbf{81.2}} & \multicolumn{1}{c}{\textbf{68.4}} & \multicolumn{1}{c}{\textbf{\underline{74.3}}} & \multicolumn{1}{c}{\textbf{838}} & \multicolumn{1}{c}{\textbf{7062}} & \multicolumn{1}{c}{\textbf{194}} & \multicolumn{1}{c}{\textbf{387}} \\\bottomrule 
\end{tabular}
}
}
\label{tab:experimentalResultsRQ2}
\vspace{-2mm}
\end{table}

Figures~\ref{fig:oracleOverlapping} and~\ref{fig:operationOracleOverlapping} show the overlapping between \approach and \baseline in terms of oracles detected (recall), grouped by oracle type and API operation, respectively. The numbers in parentheses in each tag indicate the number of possible oracles to be detected for each oracle type or API operation. Out of the 1167 oracles detected, 374 (32\%) were identified exclusively by \approach, 329 (28.2\%) only by \baseline, and 464 (39.8\%) by both approaches. Looking at Figure~\ref{fig:oracleOverlapping}, we can see that both approaches achieve similar performance in certain oracles (e.g., \texttt{string\_is\_numeric} or \texttt{number\_max\_value}).
In these cases, \approach is more cost-effective, as it only requires access to the API specification, unlike \baseline, which needs a diverse test suite to verify API behavior. However, when looking at the performance differences, it is clear that both approaches are complementary. For example, \approach can precisely infer enum values from descriptions (\texttt{string\_specific\_values}) or detect domain-specific minimum values (e.g., -90 for \texttt{latitude}). The correct inference of these oracles for \baseline may be hard if the test suite used as input is not diverse enough (e.g., it does not include a request with a latitude value of -90). On the other hand, there are some types of oracles that only \baseline detected. This is explained by the fact that some oracles simply cannot be inferred from the specification, since they may not be explicitly stated in the OAS. Instead, they require the execution of the API to find such patterns, for instance, detecting a boolean field always being true (\texttt{boolean\_always\_true}) or the maximum size of an array (\texttt{array\_max\_size}).

When analyzing the performance of \approach and \baseline per API operation (Figure~\ref{fig:operationOracleOverlapping}), the trend is similar to the one observed in the previous analysis. Although the overall recall of \approach is higher than that of \baseline, the latter detected more oracles in 10 out of 17 operations. This obviously comes at the cost of (i)~executing the API, and (ii)~being less precise, i.e., generating more false positives, as illustrated in  Table~\ref{tab:experimentalResultsRQ2}.

\begin{tcolorbox}[colback=LightGray, title=Answer to \RQ2: Static and dynamic test oracle generation effectiveness, colframe=black, left=1mm, right=1mm, top=1mm, bottom=1mm]
Overall, \approach outperforms \baseline in terms of precision, recall and F1-Score. However, their overlapping in terms of generated oracles shows that both approaches are complementary, and that \approach can generate a significant percentage of the test oracles without previously executing the API.
\end{tcolorbox}

\subsection{Experiment 2: Artificial Fault Detection}
This experiment aims to answer \RQ3 by comparing the effectiveness of the oracles generated by \approach and AGORA+ in detecting failures caused by artificially seeded faults.

\subsubsection{Experimental Setup}
Next, we describe the setup for this experiment, detailing the techniques evaluated, the test oracles implemented, the test case selection criteria, the mutant generation process and the metrics used to measure performance.

\paragraph{Techniques}
We evaluated \approach, AGORA+, and their combination in detecting API failures caused by artificial faults. We distinguish between the results of \baseline and AGORA+ (i.e., leveraging also binary oracles). We also report the failures that could be detected with all unary oracles of the ground-truth dataset, i.e., an upper bound for both \approach and \baseline.

\paragraph{Test Oracles}
For each API operation of Experiment 1, we selected the valid test oracles generated by each approach (i.e., confirmed as true positives) which were automatically transformed into executable assertions using \postmanTool~\cite{Alonso2025AGORATosem}.

\paragraph{Test Cases}
For each API operation, we randomly selected 1k API requests and responses from the set of 10k previously used (see Section~\ref{sec:exp1-setup}) meeting the following constraints: (i) they were not part of the 50-request set used as input for AGORA+; (ii) they contained at least one result item (since we cannot apply mutation operators on empty arrays); and (iii) they revealed no failures (since mutation testing requires a green test suite).
% Some failure-revealing test oracles (17 oracles from 6 APIs) exposed such a high number of failures in the unmutated API responses that they had to be discarded, despite being true positives.

\paragraph{Mutants}
Since we do not have access to the source code of the APIs under test, we cannot apply traditional mutation testing techniques. Instead, we used a \emph{black-box} approach by mutating directly the API responses. This is the same approach used by the authors of AGORA+~\cite{Alonso2023AGORA,Alonso2025AGORATosem} to evaluate the effectiveness of their approach.
% To mutate API responses, we systematically seeded \emph{errors} in API responses using JSONMutator~\cite{jsonMutator}, an open-source mutation tool that applies different mutation operators on JSON data, e.g., removing an array item. This approach differs from traditional mutation testing, where \emph{defects} are seeded in the source code of the program under test.

We used JSONMutator~\cite{jsonMutator} to introduce a \emph{single error} in each API response, simulating a \emph{failure} that could be caused by a \emph{fault} in the API.
JSONMutator is configured to apply mutation operators that result in syntactically valid mutants, i.e., conform to the API specification. Syntactically invalid mutants that would result in violations of the API specification (e.g., adding a new property to a JSON object) can be detected by existing approaches and therefore are out of the scope of both \approach and AGORA+.  Similarly, the mutation operators that convert response fields into null values are disabled, since null values can be easily detected as violations of the \texttt{nullable} property of OAS. The mutation operators applied in this context include modifications to boolean, number, and string values (e.g., by modifying or replacing values) as well as changes to array values (e.g., by removing elements or altering their order). In total, 12 different mutation operators are applied. All the mutations result in a distinguishable change in the API response and therefore there were no equivalent mutants~\cite{PAPADAKIS2019Mutation}. Our supplementary material contains a detailed list of all the mutation operators applied~\cite{satoriRepository}.

\paragraph{Metrics}
For each mutated API response of the 1k test cases used, we ran the assertions and marked the failure as detected if at least one of the test assertions failed. Then, we computed the failure detection ratio (FDR) achieved by the approach on the test suite. We repeated the mutation process 100 times
% per approach (\approach, AGORA) and for the ground truth
to minimize the effect of randomness, computing the average percentage of failures detected. In total, the results are based on 1.7M seeded errors: 17 operations $\times$ 1k API responses $\times$ 100 repetitions.

\subsubsection{Experimental Results}
Table~\ref{tab:resultsRQ3} shows the number of assertions generated (i.e., true positive oracles) and the FDR achieved by \approach, AGORA+ (unary, binary and combined), and the combination of both. \approach achieved an average FDR of 31.1\%, ranging from 14.7\% to 56.8\%. \baseline achieved an average FDR of 38.5\%, ranging from 11.5\% to 68.1\%. The binary oracles of AGORA+ increased FDR an average of 12.5\%, leading to an average FDR for AGORA+ of 51\%. The combination of both \approach and AGORA+ achieved a FDR of 55\%, ranging from 20.6\% to 92.8\%. In terms of assertions, \approach generated an average of 47.9 per API operation, more than \baseline (39.4) and less than AGORA+ (66.9).

\begin{table}[t]
\centering
\caption{\RQ3: \# assertions (A) and \% failure detection ratio (FDR) per API operation and overall by each approach.}
% \vspace{-3mm}
\resizebox{1.0\linewidth}{!}{
{
\Huge
\setlength{\tabcolsep}{3pt}
\begin{tabular}{l|cc|cc|cc|cc|cc}
\toprule
\multicolumn{1}{l}{} & \multicolumn{2}{c}{\textbf{\baseline}} & \multicolumn{2}{c}{\textbf{AGORA+ bin.}} & \multicolumn{2}{c}{\textbf{AGORA+}} & \multicolumn{2}{c}{\textbf{\approach}} & \multicolumn{2}{c}{\textbf{Both}} \\
\cmidrule(l){2-3} \cmidrule(l){4-5} \cmidrule(l){6-7} \cmidrule(l){8-9} \cmidrule(l){10-11} 
\multicolumn{1}{l}{\textbf{API -   Operation}}  & \multicolumn{1}{c}{\textbf{\#A}} & \multicolumn{1}{c}{\textbf{FDR}} & \multicolumn{1}{c}{\textbf{\#A}} & \multicolumn{1}{c}{\textbf{FDR}} & \multicolumn{1}{c}{\textbf{\#A}} & \multicolumn{1}{c}{\textbf{FDR}} & \multicolumn{1}{c}{\textbf{\#A}} & \multicolumn{1}{c}{\textbf{FDR}} & \multicolumn{1}{c}{\textbf{\#A}} & \multicolumn{1}{c}{\textbf{FDR}} \\
\midrule
AmadeusHotel & 47 & 56.4 & 22 & 3.8 & 69 & 60.2 & 70 & 48.5 & 107 & 66.9 \\ 
Deutschebahn & 32 & 19 & 6 & 0.8 & 38 & 19.8 & 53 & 16.8 & 73 & 26.5 \\ 
DHL & 10 & 45.6 & 3 & 2.8 & 13 & 48.3 & 10 & 34.8 & 19 & 51.2 \\ 
FDIC & 107 & 43.8 & 30 & 3.1 & 137 & 46.9 & 114 & 27.4 & 187 & 52.9 \\ 
GitHub-createRepo & 75 & 34.9 & 117 & 57.9 & 192 & 92.8 & 92 & 39.5 & 226 & 92.8 \\ 
GitHub-getRepos & 69 & 40.1 & 61 & 24.7 & 130 & 64.9 & 72 & 37.8 & 143 & 65.5 \\ 
GitLab-getBadges & 3 & 30.4 & 0 & 0 & 3 & 30.4 & 5 & 35.8 & 6 & 49.4 \\ 
GitLab-projectJobs & 42 & 25.5 & 11 & 11.6 & 53 & 37.2 & 48 & 23.9 & 83 & 39.8 \\ 
Marvel & 40 & 30.3 & 16 & 6.3 & 56 & 36.6 & 31 & 19 & 69 & 39.3 \\ 
OMDB-byIdTitle & 14 & 33.8 & 1 & 2.4 & 15 & 36.2 & 9 & 17.7 & 16 & 38 \\ 
OMDB-bySearch & 4 & 18.6 & 1 & 2 & 5 & 20.6 & 2 & 14.7 & 5 & 20.6 \\ 
Spotify-playlist & 18 & 46.7 & 22 & 46.1 & 40 & 92.8 & 15 & 28.9 & 46 & 92.8 \\ 
Spotify-albumTracks & 23 & 65.6 & 19 & 1.6 & 42 & 67.2 & 21 & 53.9 & 48 & 68.4 \\ 
Spotify-artistAlbums & 22 & 68.1 & 21 & 7.4 & 43 & 75.6 & 19 & 56.8 & 48 & 78.3 \\ 
Vimeo & 104 & 23.2 & 95 & 23.9 & 199 & 47.1 & 111 & 20.5 & 255 & 51.6 \\ 
Yelp & 7 & 11.5 & 5 & 12.1 & 12 & 23.6 & 17 & 15 & 24 & 31 \\ 
YouTube & 53 & 60.8 & 37 & 5.3 & 90 & 66.1 & 125 & 37.3 & 183 & 70.7 \\ 
\midrule
\multicolumn{1}{l}{\textbf{TOTAL}} & \textbf{ 670 } & \textbf{ 38.5 } & \textbf{ 467 } & \textbf{ 12.5 } & \textbf{ 1137 } & \textbf{ 51 } & \textbf{ 814 } & \textbf{ 31.1 } & \textbf{ 1538 } & \textbf{ 55 } \\ 
\bottomrule
\end{tabular}
}
}
\label{tab:resultsRQ3}
\vspace{-4mm}
\end{table}

\begin{figure}[t]
  \centering
  \includegraphics[width=1.0\linewidth]{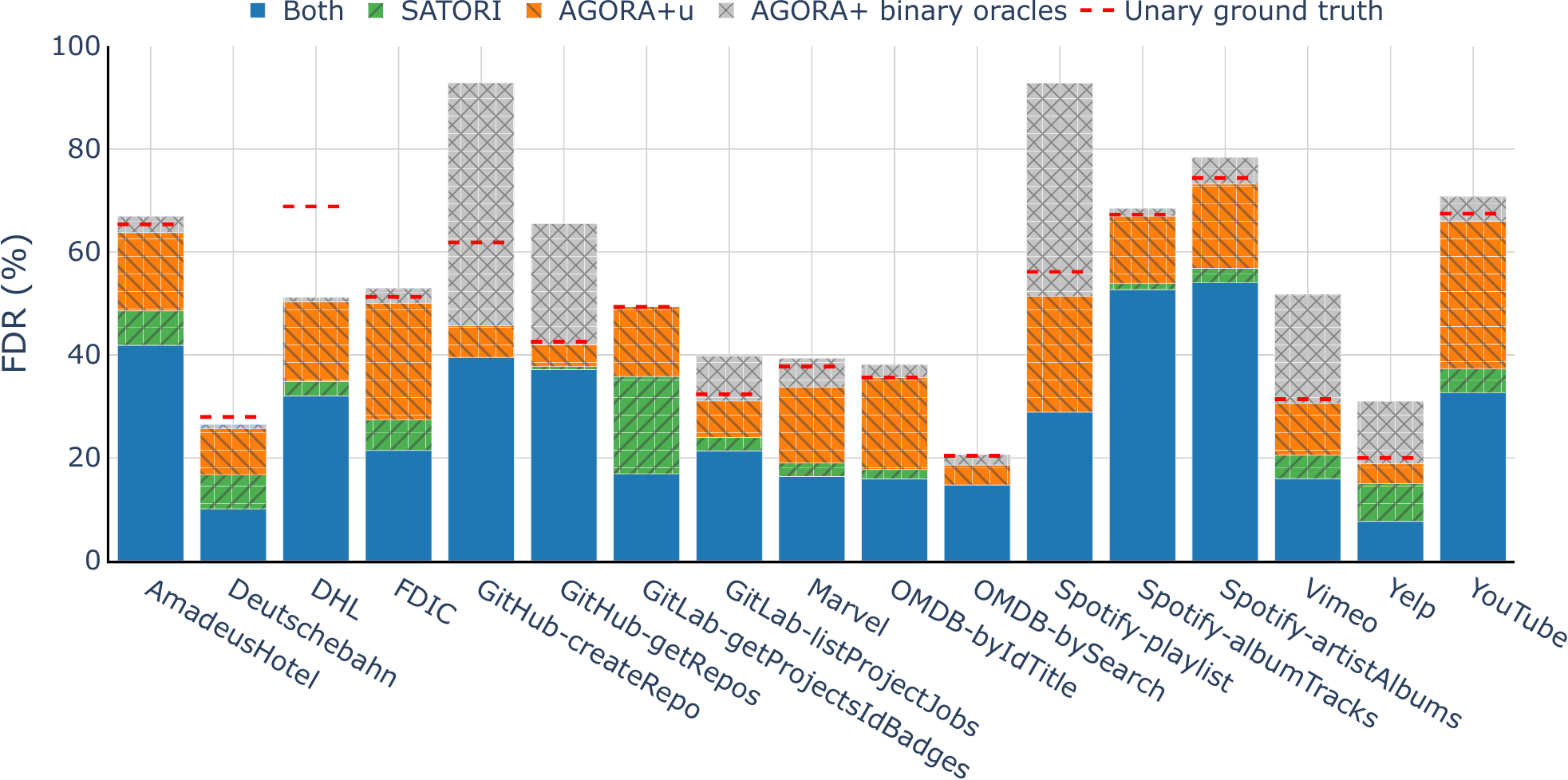}
  \vspace{-4mm}
  \caption{FDR overlapping between \approach and AGORA+.}
  \label{fig:fdrOverlapping}
  \vspace{-4mm}
\end{figure}

While \baseline achieved a higher FDR than \approach, two things are worth noting. First, \approach uncovered 80.8\% of the failures detected by \baseline (and 61\% of the failures detected by AGORA+) without needing to execute the API, which represents a significant advantage in terms of cost-effectiveness. Second, \approach managed to uncover new failures not detected by \baseline, as shown in Figure~\ref{fig:fdrOverlapping}, which represents the overlap (blue) of the FDR between \approach (green) and AGORA+ (orange), as well as the FDR achieved by the binary oracles (gray) and the optimal scenario of the ground-truth oracles (red). As illustrated, \approach detected unique failures in 14 out of the 17 API operations. This means that \approach can be used to complement \baseline, as it can detect failures that \baseline cannot, and vice versa. The combination of both approaches achieved an FDR of 55\%, which is significantly higher than the FDR of either approach alone.

Figure~\ref{fig:fdrOverlapping} also provides interesting insights regarding the strength of the unary oracles generated by \approach and \baseline combined. In most APIs, the generated unary oracles (blue, green and orange bars) achieved an FDR very close to that achieved by the ground-truth oracles, i.e., the optimal scenario (red lines). In detail, the ground-truth oracles achieved an average FDR of 47.6\% across all API operations. The automatically generated unary oracles achieved an average FDR of 44.3\%, just 3.3\% below the ground-truth oracles. This indicates that these unary oracles are very effective at detecting the failures that they are designed to detect. Intuitively, there are some failures that are impossible to detect even with the ground-truth oracles, such as subtle modifications to string fields which do not follow any format or the mutation of a number field within a certain valid range. Detecting such failures is extremely challenging and requires domain-specific knowledge or even manual inspection.

% shows a clearer overview of the overlap of the FDR between \approach and AGORA+, as well as the optimal scenario (i.e., FDR considering the oracles from the ground-truth dataset), shown as horizontal red lines. As illustrated, while there is a significant overlap between \approach and AGORA+ (blue bars), each approach also manages to detect unique failures (orange and  bars)

% the overlapping, in terms of FDR, between SATORI and AGORA+. For the mutants that were only detected by AGORA+, we distinguish between those detected by \baseline and those that required resorting to binary oracles. \approach detected 53.1\% of all the failures detected by AGORA+ (66.1\% of those detected by \baseline), plus those failures that were detected only by \approach, which shows that an important percentage of the failures can be detected without previously executing the APIs to infer test oracles.
% The horizontal red lines in the graph indicate the FDR achieved by the ground truth, which represents the upper bound for the FDR attainable using the unary oracles currently supported by both approaches. On average, the FDR of the unary oracles falls 3.3 percentage points below that of the ground truth, highlighting that combining both approaches can yield results that closely approximate the ground truth.

The APIs of GitHub, Spotify-playlist, Vimeo, and Yelp benefited from the binary test oracles of AGORA+ (gray bars), achieving a notable boost in FDR. This is primarily due to the presence of numerous equality (e.g., \texttt{input.description == return.description}), substring (e.g., \texttt{return.name substring of return.full\_name}), and arithmetic (e.g., \texttt{return.total >= size(return.businesses[])}) comparisons in these APIs, which contribute to the inflated results.

% The boost in FDR obtained by binary test oracles (gray bars) is notable in GitHub, Spotify-playlist, Vimeo, and Yelp. This is driven by the presence of numerous equality (e.g., \texttt{input.description == return.description}),
% % (e.g., \texttt{ input.description == return.description})
% substring (e.g., \texttt{return.name substring of return.full\_name}) and
% %  invariants which compare different variables (e.g., \texttt{ input.description == return.description})
% %  (e.g., \texttt{return.name is a substring of return.full\_name}) in GitHub, Spotify, and Vimeo. Additionally,
% arithmetic (e.g., \texttt{ return.total >= size(return.businesses[])}) comparisons in these operations, which contribute to the inflated results.

The higher FDR (i.e., detecting more failures) of AGORA+ over \approach (51\% vs. 31.1\%) does not necessarily mean that AGORA+ can catch more \emph{real} bugs than \approach. Our next experiment is designed to further explore this aspect.

\begin{tcolorbox}[colback=LightGray, title=Answer to \RQ3: Artificial fault detection capability, colframe=black, left=1mm, right=1mm, top=1mm, bottom=1mm]
\approach detected 61\% of the failures detected by AGORA+ without previously executing the API under test, with an FDR ranging between 14.7\% and 56.8\%. More importantly, both approaches are complementary, achieving a combined FDR of 55\%.
\end{tcolorbox}

\subsection{Experiment 3: Real Fault Detection}
\label{sec:faults}

This experiment addresses \RQ4 by comparing the effectiveness of \approach and AGORA+ in detecting failures caused by real faults.

\subsubsection{Experimental Setup}
We compared the performance of \approach and AGORA+ in detecting real failures by converting the valid test oracles reported by each approach into executable assertions using PostmanAssertify. These assertions were then executed on the original dataset of 10k API requests from Experiment 1. Violations of these test oracles revealed real bugs.
For AGORA+, we repeated the experiment 10 times, each using a different random subset of 50 requests for invariant detection, and report the number of executions in which AGORA+ identified each bug.
% This is hard to understand:
% AGORA+ can also detect bugs when an invariant exposes inconsistencies. For example, in the Amadeus Hotel API, the invariant \texttt{return.code one of \{MISCELLANEOUS, SERVICE\_CHARGE, TOTAL\_TAX\}} indicated inconsistencies in the \texttt{code} response field, which, according to the specification, should contain country codes.

\subsubsection{Experimental Results}

Table~\ref{tab:realFaults} shows all bugs found among \approach and AGORA+. \approach found 18 bugs, while AGORA+ found 13 (although only 7 of them were consistently detected in all executions). In total, both approaches detected 22 unique bugs across 8 APIs; 13 of these bugs (bugs 1, 5, 6, 7, 8, 13, 14, 15, 16, 17, 18, 19, and 20) were also detected by the authors of AGORA+~\cite{Alonso2023AGORA,Alonso2025AGORATosem}; 2 of the new bugs found only by \approach (bugs 21 and 22) have been confirmed by developers. Our supplementary material provides detailed bug descriptions, replication videos, and anonymized screenshots of reports and developer responses~\cite{satoriRepository}. These bugs are grouped into the following four categories:

\begin{table}[]
\centering
\caption{\RQ4: Faults detected by \approach and AGORA+.}
\vspace{-2mm}
\resizebox{0.82\linewidth}{!}{
\begin{tabular}{cclcc}
\toprule
\textbf{\#Bug} & \textbf{Cat.} & \textbf{API-Operation} & \textbf{AGORA+} & \textbf{\approach} \\
\midrule1 & 1 & AmadeusHotel & 1 & \ding{51} \\
2 & 2 & Deutschebahn & 10 & \ding{51} \\
3 & 1 & Deutschebahn & - & \ding{51} \\
4 & 1 & Deutschebahn & - & \ding{51} \\
5 & 1 & Deutschebahn & 1 & \ding{51} \\
6 & 2 & FDIC & 10 & \ding{51} \\
7 & 2 & FDIC & 10 & \ding{51} \\
8 & 2 & FDIC & 10 & \ding{51} \\
9 & 2 & FDIC & - & \ding{51} \\
10 & 2 & FDIC & - & \ding{51} \\
11 & 2 & FDIC & - & \ding{51} \\
12 & 2 & FDIC & - & \ding{51} \\
13 & 4 & GitHub-createRepo & 9 & - \\
14 & 1 & GitLab-projectBadges & 2 & \ding{51} \\
15 & 2 & Marvel & 1 & \ding{51} \\
16 & 3 & Marvel & 10 & - \\
17 & 1 & Marvel & 10 & - \\
18 & 1 & Marvel & 6 & - \\
19 & 1 & Marvel & - & \ding{51} \\
20 & 2 & OMDB-bySearch & 10 & \ding{51} \\
21 & 2 & Vimeo & - & \ding{51} \\
22 & 3 & Vimeo & - & \ding{51} \\
\midrule
\multicolumn{3}{l}{\textbf{Found always / Found at least once}} & 7/13 & \textbf{18/18} \\
\bottomrule
\end{tabular}
}
\label{tab:realFaults}
\vspace{-4mm}
\end{table}

\textbf{Category 1: Invalid String Formats.} These bugs occur when a string response field is expected to follow a specific format (e.g., country codes, numbers, or timestamps), but the API returns values that deviate from this format. For example, in the ``projectBadges'' operation of the GitLab API (bug 14), \approach and AGORA+ identified instances where the \texttt{rendered\_image\_url} response field, expected to be a URL, instead returned invalid URLs containing whitespaces.

% \vspace{0.25cm}
% \noindent
\textbf{Category 2: Invalid Enum Values.} These bugs are found when a response field is expected to hold specific string values (often outlined in the specification), but the API returns either undocumented values (bugs 9, 10, 11, 12, 15, 20, and 21) or entirely different ones (bugs 2, 6, 7, and 8). For instance, in the FDIC specification, the \texttt{CONSERVE} and \texttt{LAW\_SASSER\_FLG} fields are expected to use numerical flags (``1'' or ``0''), yet the API returns ``Y'' or ``N'' (bugs 6 and 7). Similarly, in the Vimeo API (bug 21), while the specification lists 14 valid values for the account field, the API also returns an undocumented value (``custom''). The Vimeo API providers confirmed this inconsistency and created an internal issue to fix it.

% \vspace{0.25cm}
% \noindent
\textbf{Category 3: Numerical Constraints.} These bugs occur when a numerical response field does not comply with expected constraints, such as minimum or maximum values. For example, in the Vimeo API, the \texttt{field\_of\_view} response field should range from 30 to 90. However, \approach detected three videos with values exceeding the upper limit (bug 22). This bug has been confirmed by the API providers, and they have created an internal issue to update the API documentation.

% \vspace{0.25cm}
% \noindent
\textbf{Category 4: Binary Test Oracles.} \sloppy{These bugs occur when a test oracle involving two variables is violated, making them detectable only by AGORA+. The only bug of this category (bug 13) was found in the ``GitHub-createRepo'' operation, where the violation of the invariant \texttt{input.license\_template==return.license.key} revealed 15 cases of repositories being created with incorrect licenses.}

\begin{tcolorbox}[colback=LightGray, title=Answer to \RQ4: Real fault detection capability, colframe=black, left=1mm, right=1mm, top=1mm, bottom=1mm]
\approach effectively detected 18 real bugs in 7 APIs.
\end{tcolorbox}

\subsection{Cost-Effectiveness Analysis}
\label{sec:cost-effectiveness}
Our last RQ investigates the monetary cost of using \approach to automatically generate test oracles for REST APIs and find bugs in them.

\begin{comment}
Things to mention:
- SATORI is a one-time cost, in the sense that it needs to be executed only on the API specification once, and then the oracles generated can be reused for all subsequent API calls. It does require using the LLM to analyze each API response.
- We computed the cost of using GPT4o with SATORI, computing the total number of input and output tokens and multiplying them by the cost of using the LLM. We considered 17 operations from 12 APIs, including 1,816 response fields (therefore 1,816 LLM API calls), which led to the generation of 10,646 test oracles.
- The total cost of this process with GPT4o was 5$.
- We found 18 bugs in total, therefore the cost per bug is 0.28$.
- We wanted to analyze whether we could actually save this cost by using a free, open-source LLM. We chose the best-performing one, i.e., Qwen2.5-32B.
- SATORI equipped with Qwen2.5-32B found 17 bugs, all except bug 10 in the FDIC API.
- Conclusion.
\end{comment}

One of the main advantages of \approach is its cost-effectiveness. Unlike what intuitively might be expected, \approach does not rely on LLMs to analyze API responses, but rather the API specification. This means that \approach can be executed once for each API response field from which one would desire to extract test oracles, and then the generated oracles can be reused for all subsequent API calls. Following this approach, we computed the marginal inference cost of using \approach with GPT-4o, the LLM that achieved the best performance in our experiments. As we considered 1,816 API response fields, we made 1,816 calls to the OpenAI API, which resulted in 716,529 input tokens and 101,949 output tokens. According to the OpenAI pricing at the time of performing the calls,\footnote{\$5 per 1M input tokens (\$2.5 if cached), \$20 per 1M output tokens~\cite{openai-pricing}.} this resulted in a total of \$5.11. As we found 18 bugs in total, the cost per bug is \$0.28.

We also analyzed whether this cost could be reduced or avoided by using a free, open-source LLM executed locally. We selected the best-performing one from \RQ1, Qwen2.5-32B. The oracles generated by \approach equipped with Qwen2.5-32B successfully found 17 bugs, missing only bug 10 in the FDIC API. This indicates that \approach is highly effective even when relying on open-source, smaller language models. However, we note that using models like Qwen2.5-32B locally requires significant computational infrastructure (e.g., high-end GPUs), and thus their actual cost depends on the availability of such infrastructure and the volume of reuse.

\begin{tcolorbox}[colback=LightGray, title=Answer to \RQ5: Cost-effectiveness analysis, colframe=black, left=1mm, right=1mm, top=1mm, bottom=1mm]
In total, \approach with GPT-4o found 18 bugs for \$5.11 (\$0.28 per bug). \approach with Qwen2.5-32B could find 17 of these bugs.
\end{tcolorbox}

\section{Threats to Validity}
\label{sec:threats}

We discuss the potential threats to the validity of our results, along with the actions taken to mitigate them.

% \vspace{0.25cm}
% \noindent
\textbf{Internal validity}. \emph{Are there factors that might affect the results of our evaluation?}
For our experiments, we used the OAS documents provided in the AGORA+ supplementary material~\cite{datasetAGORA}. 
% For the remaining APIs, we relied on the official OAS when available. In the case of Deutschebahn and Vimeo, which lack official OAS, we sourced their specifications from the APIs.guru OAS repository. 
In all cases, we updated the OAS to reflect the latest version of the web documentation.
% When possible, we used the publicly available API specifications. However, the specifications of the OMDb and Yelp APIs were unavailable, so we used manually generated specifications used by other authors in previous publications, updating them according to the latest version of the web documentation.
It is possible that these specifications have errors and deviate from the documented API behavior. To mitigate this threat, the specification files were reviewed by at least two authors.

The effectiveness of AGORA+ largely depends on the diversity of the input test suite. For a fair evaluation, we followed the same approach by the authors and used their same test suites~\cite{datasetAGORA}, leveraging the same 10 sets of 50 randomly generated request-response pairs and computing averages across the 10 runs.
% In the remaining APIs, to maximize input diversity, we followed the same approach by the authors and manually selected a set of varied test inputs for each parameter based on an analysis of the documentation. The division of this test suite into random sets may also affect the results. To mitigate this threat, we performed this division 10 times, computing the average performance between runs.
To address the potential variability of \approach, we use the default settings for all models and a temperature of 0 (greedy decoding), making the outputs of the models mostly deterministic.

Manually creating the ground truth of test oracles for all the APIs may be affected by human biases or errors. To mitigate this,
% a subset of \textcolor{red}{XXX} oracles were analyzed by two authors and no conflicts were found. Moreover, API providers were consulted in case of doubt, and the dataset is publicly available for further review~\cite{satoriRepository}.
we carefully analyzed the API specification for each response field labeled and contacted the API providers in case of doubts or discrepancies. Our supplementary material contains evidence of the questions posed to API providers and their responses, as well as the full OKAMI dataset, which is publicly available for further review~\cite{satoriRepository}.
% each classified instance of this dataset was checked by at least two authors, analyzing the API documentation, or consulting the API providers in case of discrepancies.

% \vspace{0.25cm}
% \noindent
\textbf{External validity}. \emph{To what extent can we generalize the findings of our investigation?}
We evaluated \approach using 21 different LLMs and a set of 17 operations from 12 APIs. Our conclusions may not fully generalize beyond this scope. To mitigate this threat, we selected LLMs of varying sizes and vendors, along with a set of widely-used APIs spanning diverse domains and sizes, and used in related studies.

The test oracles supported by \approach may not generalize beyond the selected APIs. We minimized this threat by basing these oracles on the unary invariants supported by AGORA+, derived from a systematic analysis of 40 real-world APIs (702 operations) from diverse domains~\cite{alonso22TSE}. However, we note that this list of test oracles is not exhaustive, and \approach is designed to be easily extended with additional oracles.

\section{Conclusions and Future Work}
\label{sec:conclusions}

This paper introduces \approach, a black-box static approach for generating test oracles for REST APIs from their specification using LLMs. \approach analyzes the response fields of an API operation, providing them as inputs to a target LLM, which infers a set of pre-defined test oracles. \approach then converts these inferred oracles into a machine-readable format compatible with an extended version of \postmanTool, a tool that transforms the oracles into executable Postman assertions. This integration makes \approach readily applicable for practical use.

Evaluation results on a set of 17 operations from 12 industrial APIs show that \approach can generate hundreds of valid test oracles per operation without executing the API. \approach achieved an F1-Score of 74.3\%. The differences in performance between \approach and AGORA+ reveal complementary strengths, with each approach excelling at detecting distinct types of test oracles, and their combination achieving a failure detection ratio of 55\%. \approach identified 18 bugs across 7 widely used industrial APIs, leading to documentation updates in the API of Vimeo. Operating in black-box mode, \approach can also be easily integrated with API testing tools that support OAS. As part of our future work, we intend to extend \approach to support test oracles involving multiple variables.

% This section is not explicitly needed, we can add it later in the camera-ready.
% \section{Data availability statement}
% We provide a supplementary material containing...\jcav{Complete}

\section{Acknowledgments}
% TOSEM
This work has been partially supported by grants PID2021-126227NB-C22 and
PID2021-126227NB-C21, funded by MCIN/AEI /10.13039/501100011033/FEDER, UE;
and grant TED2021-131023B-C21, funded by MCIN/AEI/10.13039/501100011033 and
by European Union “NextGenerationEU”/PRTR.
% new
This paper is also part of the project PID2024-156482NB-I00, funded by MICIU/AEI/10.13039/501100011033 and by the FSE+.
% AI4SE
This work is also supported by the Spanish Ministry of Science and Innovation under the Excellence Network AI4Software (Red2022-134647-T).

% This work has been partially supported by grants...

% \section*{Acknowledgment}

% This work has been partially supported by grants...

% \newpage

\bibliographystyle{IEEEtran}
\bibliography{bibliography}

\end{document}